\RequirePackage{fix-cm}
\documentclass[conference]{IEEEtran}

\usepackage[T1]{fontenc}
\usepackage{amsmath,amssymb}
\usepackage{graphicx}
\usepackage{xcolor}
\usepackage{booktabs}
\usepackage{array}
\usepackage{tabularx}
\usepackage[hyphens]{url}
\usepackage{cite}
\usepackage{pgfplots}
\usepgfplotslibrary{groupplots}
\pgfplotsset{compat=1.18}
\usepackage{placeins}
\usepackage{float}
\usepackage[font=footnotesize,labelfont=bf,labelsep=colon]{caption}
\floatstyle{plaintop}\restylefloat{table}
\usepackage[hidelinks]{hyperref}
\hypersetup{pdftitle={CHIP: Contrastive Hardware-Image Projection. Can OpenAI CLIP Reconstruct Images from the CPU Computational Footprint?},
  pdfauthor={Debayan Sengupta, Darsh Asher, Kaushal Mhapsekar, Bita Aslrousta, Amna Hassan, Samira Mirbagher Ajorpaz},
  pdfkeywords={side channels, performance counters, machine-learning privacy, image reconstruction, CLIP and unCLIP}}
\newcommand{\SPR}{Sapphire Rapids}

\DeclareUrlCommand\ctr{\urlstyle{tt}}

\newcommand{\chipoakoriginal}[1]{%
  \includegraphics[width=#1]{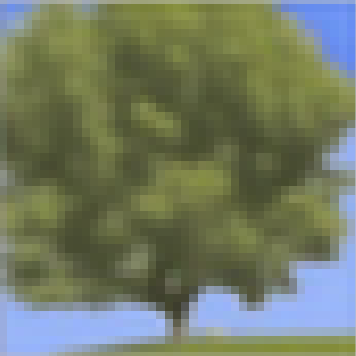}}
\newcommand{\chipoakdecoded}[1]{%
  \includegraphics[width=#1]{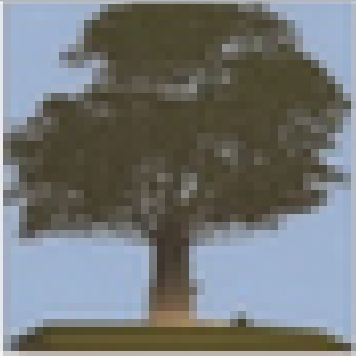}}

\newcommand{\figref}[2][]{\textbf{Figure~\ref{#2}#1}}

\IfFileExists{T1LinuxBiolinumT-TLF.fd}{}{}

\providecommand{\Description}[2][]{}
\newenvironment{acks}{\section*{Acknowledgments}}{}
\makeatletter
\def\chip@addpunct#1{\relax\ifhmode\ifnum\spacefactor>\@m\else#1\fi\fi}
\def\chip@runin#1{{\sfcode`\.=1001 \sfcode`\?=1001 \sfcode`\!=1001
  \def\@IEEEsectpunct{\chip@addpunct{.}}#1}}
\renewcommand\paragraph{\@startsection{paragraph}{4}{\parindent}%
  {4pt plus 2pt minus 1pt}{-0.5em}{\normalfont\normalsize\chip@runin}}
\makeatother

\begin{document}
\nocite{ackley1985boltzmann,akinsanya2024timing,alam2019secure,asher2026featurebleed,barlow1961,barlow1972,%
  barlow2001,batina2019csinn,carlini2024remote,chen2026uncorebleed,cook2022biggerfish,costan2016sgx,%
  das2019sok,demme2013malware,deng2009imagenet,ding2025moecho,dizani2025thor,dosovitskiy2021vit,%
  fedus2022switch,fredrikson2014pharmacogenetics,friston2007,gao2024deeptheft,gao2025iknow,gast2025counterseveillance,%
  gulmezoglu2017perfweb,hadad2026shape,hazelwood2018applied,he2016resnet,hendrycks2016gelu,hinton1986distributed,%
  hinton2006deep,hinton1993autoencoders,horvath2025barracuda,howard2019imagenette,hu2020deepsniffer,isca-v5:hua2018,%
  huegle2023power2picture,hunt2018chiron,intel2020rapl,intel2022doit,intel2022throttling,intel2022consttime,%
  intelsdm,isca-v5:gatebleed,kalyanapu2025amxgating,krizhevsky2009cifar,krizhevsky2012alexnet,lequoc2020securetf,%
  lecun1998mnist,levy1996,li2022multiexit,perfeventopen,lipp2016armageddon,lipp2021platypus,%
  liu2023llava,isca-v5:adamw,maia2022hear,maji2021leakynets,maji2024sparseleakynets,majumdar2022characterizing,%
  mehnaz2022sensitive,micay2016perfparanoid,ajorpaz2022evax,ajorpaz2020perspectron,moini2021bnn,naghibijouybari2018rendered,%
  nvidia2018renderedresponse,olshausen1996,paszke2019pytorch,pothukuchi2021maya,qiu2026keytar,qiu2023pmuspill,%
  qu2023unsafe,radford2021clip,ramesh2022unclip,rao1999,rao2021dynamicvit,schwarz2019netspectre,%
  scotti2023mindeye,shokri2017membership,shukla2023whispering,shusterman2019robust,son2026ovrwatcher,srivastava2026ghosttrace,%
  stabilityunclip,hai2025aiindex,taneja2023hotpixels,teerapittayanon2016branchynet,linuxperfsecurity,tramer2019slalom,%
  tsai2017graphene,oord2018cpc,wang2022stealthy,wang2021leakedweb,wang2024gpuzip,wang2022hertzbleed,%
  wang2004ssim,wei2018iknow,weissteiner2025teecorrelate,wolfyosys,xie2024osworld,xu2015controlled,%
  yan2020cachetelepathy,yin2022avit,yuan2025ciphersteal,yuan2022automated,yuan2021private,zhai2022scaling,%
  zhang2024timewilltell,zhang2023headset,zhang2024demystifying}

\title{\textsc{CHIP}: Contrastive Hardware--Image Projection\\
Can OpenAI CLIP Reconstruct Images from the CPU Computational Footprint?}

\author{
\IEEEauthorblockN{Debayan Sengupta, Darsh Asher, Kaushal Mhapsekar, Bita Aslrousta,\\
Amna Hassan, Samira Mirbagher Ajorpaz}
\IEEEauthorblockA{North Carolina State University\\
\{dsengup2, dkasher, kmhapse, baslrou, ahassan9, smirbag\}@ncsu.edu}
}

\maketitle
\thispagestyle{plain}\pagestyle{plain}

\begin{abstract}
Image inference on a shared CPU leaves input-dependent traces in software-visible performance counters and execution time. CHIP (Contrastive Hardware–Image Projection) trains a lightweight residual projector on public images to map these measurements into the 768-dimensional visual embedding space of a frozen OpenAI CLIP model. A frozen Stable unCLIP generator uses the projected embeddings to reconstruct recognizable color scenes for held-out inputs. The attack requires no physical probe, candidate image gallery, domain-specific generator, or shared cache, and elapsed time alone retains the signal.

We evaluate full-counter, single-counter, and timing-only measurements on Intel Haswell and Sapphire Rapids. The projected embeddings achieve 0.85 ROC AUC for held-out image identification, compared with a 0.50 baseline. Exact-image top-five retrieval reaches 24\% on CIFAR-100 and 56\% on Unsafe-Diffusion. Across both processors, reconstructions average 1.4× the baseline RGB color-palette overlap and 2.5× the baseline structural similarity.
\end{abstract}


\section{Introduction}
\label{sec:paper-introduction}

The vast
majority of artificial intelligence runs on shared cloud infrastructure. In 2024, 78\% of surveyed organizations reported using AI, up from 55\% a year earlier~\cite{hai2025aiindex}, and datacenter inference has long relied on CPUs~\cite{hazelwood2018applied}. These workloads classify images with networks such as ResNet~\cite{he2016resnet} and generate them with text-to-image models~\cite{ramesh2022unclip}; video extends both to sequences of frames. Image privacy extends beyond class labels to who appears, where, and what is happening.\looseness=-1

To keep data private from cloud providers, systems such as Slalom by Tram{\`e}r and Boneh~\cite{tramer2019slalom}, Chiron~\cite{hunt2018chiron}, and secureTF~\cite{lequoc2020securetf} run models or data in Intel SGX enclaves~\cite{costan2016sgx}, protecting them from a compromised operating system or hypervisor. Slalom and Chiron both acknowledge that SGX leaves side channels unaddressed~\cite{tramer2019slalom,hunt2018chiron}. Slalom considers side channels to be a minor DNN concern because standard DNNs apply the same operations regardless of input~\cite{tramer2019slalom}. Under this view, encryption and isolation leave little to infer from an image's computation.\looseness=-1

\begin{figure}[!t]
\centering
\includegraphics[width=\columnwidth]{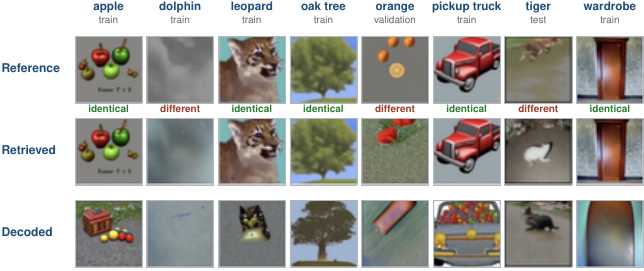}
\caption{CHIP's image reconstruction.}
\Description{Example images illustrating gallery retrieval and image generation from CHIP's estimated CLIP embeddings.}
\label{paper:fig:seen-path}
\end{figure}

In this paper, we show that this view does not hold: an SGX enclave
protects an image in memory, but not the footprint its computation
leaves in CPU measurements. For each top-row reference in \figref{paper:fig:seen-path}, CHIP's first-ranked candidate appears in the middle row and an image generated from its estimate in the bottom row. The output is an image, not a label: object category, color and coarse layout survive a few counter totals and elapsed time, without a power probe, memory-access trace or decoder trained on victim data. The figure includes marked training-set examples; every reported score uses held-out images.\looseness=-1

Previously, four assumptions made image recovery seem impractical. First, semantic recovery appeared to need a large labeled corpus pairing hardware traces with images or classes for every domain. Second, whole-inference or generation timing appeared too coarse to reveal useful content; prior image recovery relied on richer observations, such as power waveforms~\cite{wei2018iknow,maji2021leakynets} or page-access and cache-access traces~\cite{xu2015controlled,yuan2021private}. Third, image synthesis appeared to need victim data to train a decoder, such as a GAN~\cite{yuan2021private,yuan2022automated,huegle2023power2picture}. Fourth, small content-induced differences in microarchitectural activity seemed likely to vanish in system noise~\cite{das2019sok}. Three developments of the past five years challenge these assumptions.

The first development is large pretrained models with broad, shared
representations. Earlier vision models predicted fixed categories,
such as the 1,000 ImageNet
classes~\cite{deng2009imagenet,krizhevsky2012alexnet}, so each new
concept needed new labeled data~\cite{radford2021clip}. Instead, CLIP 
learns a shared image--text embedding from 400 million internet
images and captions~\cite{radford2021clip}, and later models reusing
that embedding
include the unCLIP image
generator~\cite{ramesh2022unclip}, the LLaVA visual
assistant~\cite{liu2023llava} and the MindEye fMRI
decoder~\cite{scotti2023mindeye}. Profiles of public images can replace a victim-specific labeled corpus, while a pretrained generator removes the need for a victim-domain decoder, making reconstruction from execution a practical question. CLIP's embedding has thus become a
shared interface between models: MindEye maps brain activity into
this space to reconstruct the images a person
views, but no prior work has done the same for CPU computation.\looseness=-1

For an attacker, these developments replace learning image semantics and synthesis from hardware traces with aligning execution to a shared embedding. Like MindEye's brain-activity mapping, this alignment treats execution as another view of an image. We take the first step with the unprivileged counters and timing available to any co-resident user, and show that architectural events reflect the semantics of the computation they count. This finding may extend to microarchitectural channels beyond our three variants and enable applications such as using the CPU as a root of trust for AI-agent monitoring.

The second development is AI hardware and software optimization. To save energy and latency, processors power-gate idle accelerator units, while models skip tokens, layers or experts depending on input~\cite{rao2021dynamicvit,yin2022avit,teerapittayanon2016branchynet,li2022multiexit,fedus2022switch}. These optimizations make execution data dependent and can amplify differences. Waking a power-gated Intel AMX unit can take up to 20,000 cycles, revealing early exits and mixture-of-experts routing; scheduling an AMX instruction just before power-gating can turn a 200-cycle cache miss into an approximately 11,000-cycle gap that survives reduced-resolution timers~\cite{kalyanapu2025amxgating,isca-v5:gatebleed}. Dynamic frequency scaling converts data-dependent power into observable time~\cite{wang2022hertzbleed}. Such amplification challenges the expectation that content effects vanish in noise.

\begin{figure}[!t]
\centering
\includegraphics[width=\columnwidth]{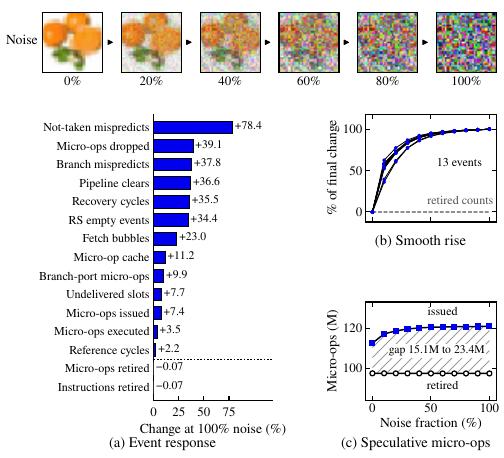}
\caption{Noise changes events and issued/retired micro-ops even when end to end the inference time is constant.}
\Description{An image is mixed with noise at a fixed shape. A bar chart ranks fifteen hardware events by their change at full noise, and a line plot shows issued micro-operations rising while retired micro-operations stay constant.}
\label{fig:noise-ladder}
\end{figure}

The third development is learned specialization. A Switch Transformer activates one expert per token~\cite{fedus2022switch}, while larger, better-trained models learn representations that generalize further~\cite{zhai2022scaling}. Different images elicit activation patterns and computational footprints. Training makes selected CNN and ResNet-18 channels class selective; we trace how content-specific activations reach executed instructions, micro-ops and time. These differences challenge the view that whole-inference measurements are too coarse to carry image content.\looseness=-1

Together, these developments undercut the four assumptions: a pretrained embedding and generator replace victim-labeled data and a victim-trained decoder, amplification keeps content effects above noise, and specialization makes whole-inference measurements content dependent. We hypothesize that a few counter totals and elapsed time carry enough image content for reconstruction. If this holds, there are three implications for processor design. First, processor design must consider what whole-inference instruction and micro-op counts reveal about learned content, not only which addresses or branches depend on secret bits. Second, enclaves and memory encryption protect data in memory, not its computational signal, calling for protections that hide instruction and micro-op counts as well as data accesses. Third, this channel could support unsafe-content or parental controls, or check AI agents that conceal what they process. Computer-use agents in OSWorld choose each action from a desktop screenshot and recent actions~\cite{xie2024osworld}, making execution's image semantics relevant to their oversight, privacy and hardware design.

Image semantics encompass objects and scenes, their attributes and arrangement, visual style, and mood or feeling. Semantics differ from pixels: two photographs can share meaning while differing in every pixel. We measure what execution reveals through exact-image identity, color and coarse structure, object class, and safety attributes.

Recovering image semantics from hardware is difficult: each measurement summarizes an entire inference. A few counter totals and elapsed time vary only slightly across images, and regressing $224\times224\times3$ pixel values is ill-posed. We ask whether this variation encodes enough image-specific information to locate the image in a pretrained representation for retrieval or synthesis; %
to our knowledge, no prior work maps CPU execution into CLIP space.\looseness=-1

We introduce CHIP (\emph{Contrastive Hardware--Image Projection}), which estimates 768 CLIP embedding coordinates from aggregate software-visible counters or elapsed time rather than predicting $150{,}528$ pixels. The representation supports reconstruction, exact-image retrieval, identification, and class or safety inference, beyond recovering labels, membership or model properties. CLIP and the publicly released Stable unCLIP generator~\cite{stabilityunclip}, conditioned on the same ViT-L/14 embedding, remain frozen. Synthesis thus needs neither a domain-trained generator nor private images to train one; the same representation serves CIFAR-100~\cite{krizhevsky2009cifar}, Unsafe-Diffusion~\cite{qu2023unsafe} and both victim architectures.

Only a residual multilayer perceptron is trained, pairing each public image's measurements with its CLIP embedding. The pretrained encoder and generator supply visual structure; the learned mapping supplies execution information. We test how input values affect executed instructions, micro-ops and timing, linking recovered content to architectural causes.

We evaluate CLIP ViT-L/14 on Intel Haswell and Sapphire Rapids and study ResNet-18 retrieval. Offline profiling reaches 24\% exact-image top-five retrieval on CIFAR-100 ($14.5\times$ chance) and 56\% on Unsafe-Diffusion ($9\times$ chance); candidate identification reaches 0.85 ROC AUC against 0.50 chance. Reconstructions without a candidate set close 75\%--84\% of the palette gap between shuffled-trace and true-embedding generation on both Intel CPUs. \figref{paper:fig:seen-path} shows both attack outputs. ResNet-18, which has no token pruning, ranks the exact image first for six of nine displayed queries in a 2,000-image candidate set, so the leakage is not specific to pruning.
\label{sec:attack-overview}

The signal persists across tested conditions. Except for marked training-set illustrations (\figref{paper:fig:seen-path}), reconstruction and retrieval use held-out images and shuffled-trace controls pairing measurements with wrong images. Under offline profiling, controls yield 1.4\%--1.8\% top-five versus 24.2\% with true traces. Processor generations eight years apart reach 29.9\% and 30.1\% top-five on a common 75-image candidate set. One measurement per image retains about 95\% of the ten-measurement score; noise equal to the full between-image spread still leaves $6.6\times$ chance retrieval; and unaveraged single traces in cross-core tests reach 27.6\% top-five versus 28.0\% with averaging.

What makes execution distinguishable? Learning organizes representations around regularities. Boltzmann and deep-belief models express learned preference through an energy landscape~\cite{ackley1985boltzmann,hinton2006deep}; economical codes~\cite{hinton1993autoencoders,olshausen1996,barlow1961,barlow1972,barlow2001}, residual signaling~\cite{rao1999,friston2007}, and neural signaling costs~\cite{levy1996} connect representation to computational economy. These theories motivate examining activation magnitudes, zeros and bit patterns, but CPU leakage requires establishing their links to executed instructions, micro-ops and observable time.

We link leakage to representation learning: activations vary with content, shaping the processor's observable counters and timing. Two controlled tests examine this link. With weights and input shape fixed, mixing noise into an image changes speculative execution, including mispredicted branches and issued micro-ops, while retired instructions and micro-ops stay the same (\figref{fig:noise-ladder}). The smooth rise is unexpected: identical layers and tensor shapes might suggest constant counters or irregular changes due to noise. Training makes channel firing depend on class (\figref{fig:selectivity}). Our CLIP ViT-L/14~\cite{radford2021clip} uses QuickGELU, not ReLU, so ReLU-induced sparsity is not required for the leakage we observe.

Unlike Yuan et al.~\cite{yuan2021private,yuan2022automated}, we use unCLIP as is, without training the decoder. The recovered semantics therefore come from the measured computation rather than from a decoder fitted to the victim's domain.\looseness=-1

We found that the fixed cycle counters \path{IA32_FIXED_CTR1/2} remain unsuppressed inside SGX enclaves. With privileged monitoring, UncoreBleed also observes unsuppressed uncore events: CHA \path{TOR_INSERT}, M2M \path{TAG_HIT}, and iMC \path{ACT_COUNT} and \path{CAS_COUNT}~\cite{chen2026uncorebleed}. Opt-out enclave entry suppresses retired-instruction and core-programmable counters, but the fixed cycle counters keep counting~\cite{intelsdm}; our Gramine-SGX~\cite{tsai2017graphene} check confirms that programmable counts stay nearly constant while cycles and elapsed time survive. Like the RAPL energy interface that PLATYPUS~\cite{lipp2021platypus} used to leak secrets from SGX enclaves, this counter is exposed for performance monitoring yet reveals what the enclave computes. Restricted to cycles and time under offline profiling, ordinary Haswell inference traces reach 16.5\% top-five retrieval against 1.67\% chance. An attacker controlling the OS, the standard SGX adversary, can observe image-dependent cycle counts even when the enclave protects the victim's memory.

The execution-content association motivates defenses. Hardware performance-counter detectors recognize malicious activity~\cite{demme2013malware,ajorpaz2020perspectron,ajorpaz2022evax}; CHIP instead estimates processed content without relying solely on application reports, potentially supporting parental controls on shared devices. Safety classification reaches 66.2\% on one split and 60.2\% under five-fold evaluation against 50\% chance. These findings motivate hardware-assisted safe/unsafe monitoring when the application is untrusted, alongside workload-privacy protections.

\noindent\textbf{Contributions.} This paper makes three contributions:
\begin{enumerate}\setlength{\itemsep}{1pt}\setlength{\parskip}{0pt}
\item \textbf{From CPU measurements to image content.} We map whole-inference measurements into CLIP's pretrained im\-age--text representation, extending prior class-label attacks to color-image reconstruction. This avoids learning $224 \times 224 \times 3$ pixels from scratch, supports Vision Transformers~\cite{dosovitskiy2021vit} without relying on CNN-specific ReLU or max-pooling branches, and requires no power probe.{\emergencystretch=1.5em\par}

\item \textbf{CHIP: learning the mapping with CLIP and unCLIP fixed.} A residual network aligns measurements with CLIP embeddings while preserving the frozen generator's expected representation. Training uses paired measurements and embeddings, or offline profiles from a public model copy with linear calibration and checkpoint selection on the target CPU. The shared representation supports CLIP ViT-L/14 and ResNet-18 victims.

\item \textbf{Image reconstruction, identification, retrieval, and safety inference.} We evaluate unprivileged timing (CHIP-T), permitted counters (CHIP-C), and an enclave under a compromised OS (CHIP-E) on Intel Haswell and Sapphire Rapids, across CIFAR-100 and Unsafe-Diffusion. A retired-instruction counter carries the full eight-event vector's retrieval signal; leading counters count retired instructions and micro-ops, elapsed time retains most of the signal, and ResNet-18 also exposes image identity. Finally, we discuss the multi-factor root cause, vendor-suggested mitigations and their overhead.

\end{enumerate}

\textbf{Responsible disclosure.}
We disclosed these findings promptly to Intel, who reproduced our inference results. The proposed patch disables the affected performance counters, such as \path{IA32_FIXED_CTR1/2}, particularly in enclaves; a CVE number is pending. %

\section{Background and Related Work}
\label{sec:paper-background}
\label{sec:related}

Prior work differs from CHIP in three ways: what it recovers, what it must observe, and whether it has seen its target before. Prior work recovers a label, a membership bit or a model property rather than an image; it recovers inputs only with a power sensor, a fine-grained or privileged access trace, or a generator trained on the victim's domain; and, when it exploits counters or browser timing, it recognizes only workloads profiled in advance. CHIP's signal is not confined to one counter: a retired-instruction counter (\path{IA32_FIXED_CTR0}) alone nearly matches the full vector's retrieval accuracy on Haswell, while cycle counters (\path{IA32_FIXED_CTR1/2}) and programmable counters for micro-ops, cache activity, prefetches and branches also reflect image content (\S\ref{paper:sec:measurements}). To our knowledge, CHIP is the first to map these aggregate CPU measurements into CLIP space to reconstruct images unseen during attack training. We verified unprivileged access to permitted core counters on the tested cloud nodes; Section~\ref{sec:mitigation} discusses microcode updates and other mitigations.\looseness=-1

\vspace{2pt}\noindent{\normalfont\bfseries Software-visible leakage from inference}
Software-visible attacks establish that neural-network execution leaks information, but they recover targets coarser than the image. CPU-counter and cache studies recover class-dependent distributions or class labels of CNN inputs~\cite{alam2019secure,majumdar2022characterizing,wang2022stealthy}, and inference-timing attacks recover predicted classes~\cite{shukla2023whispering,maji2024sparseleakynets,akinsanya2024timing}; a label supports neither exact-image retrieval nor synthesis. GATEBLEED~\cite{isca-v5:gatebleed}, multi-exit attacks~\cite{li2022multiexit} and GhostTrace~\cite{srivastava2026ghosttrace} expose membership or routing decisions. Membership inference~\cite{shokri2017membership} and attribute inference~\cite{fredrikson2014pharmacogenetics,mehnaz2022sensitive} concern the records a model was trained on, not the input the victim is currently computing on. FeatureBleed~\cite{asher2026featurebleed} infers private enriched features of tabular inputs and the output class of images, and Thor~\cite{dizani2025thor} recovers hidden weight sparsity from AMX timing of a single layer. Each of these attacks targets a property of the model, its training data or its decision, whereas CHIP recovers the visual content of the input.\looseness=-1

\vspace{2pt}\noindent{\normalfont\bfseries Input reconstruction through power}
Attacks that recover inputs have needed a physical power sensor, a fine-grained access trace~\cite{yuan2025ciphersteal,chen2026uncorebleed}, or a generator trained on the victim's domain. Wei et al.~\cite{wei2018iknow} use an external power probe, Leaky Nets~\cite{maji2021leakynets} a shunt and oscilloscope, and Moini et al.~\cite{moini2021bnn} on-chip voltage sensors of a remote FPGA. Media-software attacks reconstruct images from page faults~\cite{xu2015controlled}, from privileged uncore-counter traces of an enclaved libjpeg decoder at cache-line granularity~\cite{chen2026uncorebleed}, or decode cache-line and page-table traces with a variational autoencoder or generative adversarial network trained on the attacked image domain~\cite{yuan2021private,yuan2022automated,zhang2024demystifying}, and power-based reconstruction likewise trains its generator on the victim domain~\cite{huegle2023power2picture}. A domain-trained generator confines every output to that domain and requires many victim images, which are private because they are the target. These attacks therefore reconstruct only images from the domain their decoder was trained on. CHIP needs neither the sensor nor the trace nor the domain-trained decoder: it uses aggregate counters or elapsed time, and its frozen CLIP encoder and Stable unCLIP generator were never trained on victim data.\looseness=-1

\vspace{2pt}\noindent{\normalfont\bfseries Text, multimodal, and rendered-pixel leakage}
Shape and Substance~\cite{hadad2026shape} infers VLM image geometry and content density from preprocessing time and last-level-cache misses, and MoEcho~\cite{ding2025moecho} completes masked images from mixture-of-experts load footprints with 40--50\% of the pixels visible, whereas CHIP synthesizes an unseen image from aggregate execution of a vision encoder with no pixels revealed. I Know What You Said~\cite{gao2025iknow} and KeyTAR~\cite{qiu2026keytar} reconstruct text from cache observations or keystroke timing, and timing studies infer query or output attributes~\cite{carlini2024remote,zhang2024timewilltell}; CHIP instead recovers images. GPU.zip~\cite{wang2024gpuzip} and Hot Pixels~\cite{taneja2023hotpixels} steal rendered browser pixels rather than a model's input. The attacker embeds a cross-origin page, renders it repeatedly through filters of its choosing, and leaks one binarized pixel at a time through memory-bus, cache or frequency effects (0.2--6.2 pixels per second for GPU.zip), yielding a two-level bitmap that can expose rendered text but not color or scene content. CHIP controls nothing the victim computes and shares no cache or memory with it; from aggregate measurements of one inference, it reconstructs the image's objects, color and layout, retrieves the exact source, and identifies its class and content safety.

\vspace{2pt}\noindent{\normalfont\bfseries Model extraction, fingerprinting, and counter-based monitoring}
Model-extraction attacks recover layer topology or weights from cache, bus-timing or electromagnetic observations~\cite{yan2020cachetelepathy,hu2020deepsniffer,isca-v5:hua2018,batina2019csinn,maia2022hear,horvath2025barracuda,gao2024deeptheft}, whereas CHIP does not extract the model: it recovers the private image the model processes. Counter-based work stops at recognizing activity that it has seen. Malware and attack detectors~\cite{demme2013malware,das2019sok,ajorpaz2020perspectron,ajorpaz2022evax} feed microarchitectural events to a classifier; website, browser and AR/VR fingerprinting~\cite{wang2021leakedweb,shusterman2019robust,naghibijouybari2018rendered,gulmezoglu2017perfweb,zhang2023headset,son2026ovrwatcher} identifies which program or site is running, after which NVIDIA restricted GPU-counter access~\cite{nvidia2018renderedresponse}; and TEEcorrelate~\cite{weissteiner2025teecorrelate} monitors an AMD SEV-SNP guest through host counters~\cite{gast2025counterseveillance}. Each of these attacks trains a classifier on traces of the same sites, applications or activities that it later recognizes, so its output is the name of a profiled workload and never content absent from its training set. CHIP uses the same class of software-visible events to reconstruct held-out images that appear in none of its training traces. ARMageddon~\cite{lipp2016armageddon} demonstrates cache attacks on ARM and an Android kernel change restricted performance events~\cite{micay2016perfparanoid}, whereas CHIP needs no shared cache and retains a timing channel when counters are restricted.\looseness=-1

\section{Threat Model}
\label{sec:paper-threat-model}
\label{sec:threat}

Like PLATYPUS~\cite{lipp2021platypus}, CHIP considers an unprivileged co-resident attacker and a privileged attacker targeting an SGX enclave, but uses elapsed time and permitted CPU counters rather than the RAPL energy interface~\cite{intel2020rapl}. The timing-only setting follows prior remote and cross-process attacks~\cite{wang2022hertzbleed,schwarz2019netspectre,isca-v5:gatebleed}. The victim sends a private image to a CPU-hosted vision inference service and receives a private embedding or content-safety decision; the model architecture, weights, inference configuration and server memory remain private. CHIP's basic attack is black-box, requiring only observable inference boundaries and measurements, not model internals or training data. With ordinary client permissions, the attacker profiles known public images through the same API, pairing their measurements with embeddings from its public CLIP encoder. At attack time, it measures a victim request to estimate the private image's identity and semantic content. 
Physical access, power or electromagnetic probes and vulnerabilities in the service are out of scope.\looseness=-1

\textbf{CHIP-T}, the baseline, measures end-to-end request latency at the API boundary or elapsed time from an unprivileged co-located process, requiring neither monitoring permission nor shared memory or cache.
\textbf{CHIP-C} adds performance counters, which prior attacks read on AR/VR headsets~\cite{zhang2023headset} and GPUs~\cite{srivastava2026ghosttrace}. In the same-core setting, an attacker on a sibling hardware thread observes shared-core effects through permitted user-mode counts~\cite{linuxperfsecurity,perfeventopen}. On tested public-cloud nodes with \path{perf_event_paranoid}=2, an ordinary user without added capabilities could read all CHIP-C core events (instructions, cycles, micro-ops, cache and branches) as per-task user-mode counts; kernel-mode, system-wide, uncore and RAPL readings were denied. This checks counter availability, not the probe channel; \S\ref{sec:paper-methodology} describes the evaluated per-task measurements.
\textbf{CHIP-E} assumes a compromised host OS~\cite{xu2015controlled,qiu2023pmuspill}, while SGX protects enclave code and data; opt-out entry leaves fixed cycle counters and elapsed time observable~\cite{intelsdm}.

\section{CHIP}
\label{sec:paper-attack}

\subsection{Overview}
\label{sec:chip-overview}
\label{paper:sec:completeness}
\label{sec:leakage}

Before describing the algorithm, we use controlled studies to show which observations motivate the attack and what enables this semantic inference, in five steps. Each is a controlled observation about the computation and fixes one choice in the algorithm of \S\ref{sec:attack-building}; the accuracy of the resulting attack is evaluated in \S\ref{sec:paper-results}.

\paragraph{\normalfont\bfseries Step 1: Does Content Change the Counters at a Fixed Input Shape?}
\label{root:noise}
We start with the most restrictive setting: fixed weights and a fixed input shape, so layer sequence and tensor dimensions are identical for every input. If the footprint still moves, the cause can only be the values that flow through the network. For 1,000 CIFAR-10 images~\cite{krizhevsky2009cifar}, we mix each image with same-shaped random noise in eleven fractions from 0 to 1 and run AlexNet~\cite{krizhevsky2012alexnet} with weights initialized once, in PyTorch~\cite{paszke2019pytorch} with oneMKL on \SPR{} with prefetchers and SMT disabled.

The footprint does move (\figref{fig:noise-ladder}). Of 263 events, 56 track noise nearly monotonically ($|\rho|\geq0.95$). Mispredicted branches rise 37.8\%, recovery cycles 35.5\% and issued micro-ops 7.4\%, while retired instructions and micro-ops change $-0.07\%$: nearly identical retired micro-ops hide different issued micro-ops. The response is also smooth (\figref[(b)]{fig:noise-ladder}). With the same layers and tensor shapes at every step, the expectation is a constant footprint or one that zigzags with the noise sample; instead, thirteen responding events in the figure rise at every step (others noisy or constant). \textit{A footprint that varies continuously with content is matched by a continuous target, not by a fixed set of labels. This first design choice, regressing to a point in continuous space, is one reason CHIP works.}\looseness=-1

\begin{figure}[!t]
\centering
\includegraphics[width=\columnwidth]{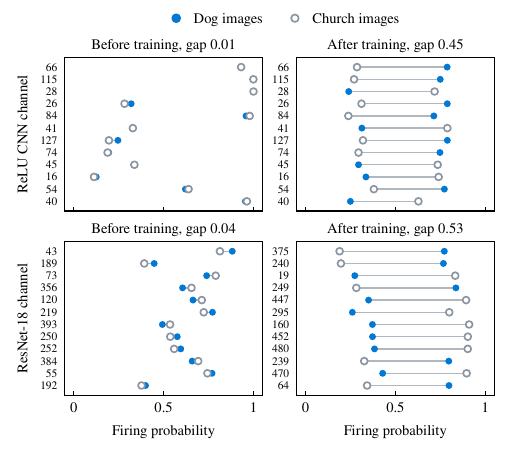}
\caption{Training splits dog and church firing.}
\Description{Before-and-after training comparisons show dog-versus-church channel firing in a CNN and ResNet-18.}
\label{fig:selectivity}
\end{figure}

\paragraph{\normalfont\bfseries Step 2: Does Learning Make Feature Firing Depend on Class?}
\label{root:selectivity}
With fixed random weights, that content is not yet organized by meaning, so we ask what training changes. We compare a four-layer CNN before and after 12 Imagenette~\cite{howard2019imagenette} epochs, and random versus ImageNet-trained~\cite{deng2009imagenet} ResNet-18~\cite{he2016resnet}, on twelve dog--church-discriminating channels selected on auxiliary images and evaluated on disjoint validation images. Firing is the nonzero fraction of a channel's spatial outputs.

\begin{figure}[!htbp]
\centering
\includegraphics[width=\columnwidth]{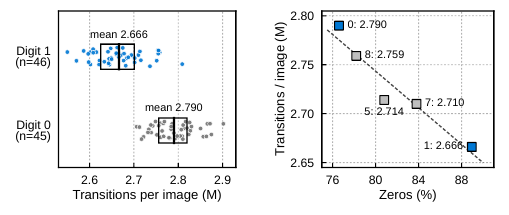}
\par\vspace{1pt}
\makebox[0.51\columnwidth][c]{\scriptsize\bfseries (a) Digits 1 and 0}\hfill
\makebox[0.47\columnwidth][c]{\scriptsize\bfseries (b) Zeros and transitions}
\caption{Inputs with more zeros cause fewer bit transitions.}
\Description{Two panels compare bit transitions for digits 1 and 0, and the relation between zero counts and bit transitions across five digits.}
\label{fig:flip-chain}
\end{figure}

Training makes neuron firing depend on class (\figref{fig:selectivity}): the mean class gap of the selected channels rises from 0.01 to 0.45 in the CNN and from 0.04 to 0.53 in ResNet-18. The randomly initialized CNN's twelve most discriminating channels give a 0.05 gap, so learning, not selection alone, makes zero patterns content-informative. This is the behavior of a learned distributed representation~\cite{hinton1986distributed}: each input is represented by a pattern of activity over many units, so which units respond to an image is the network's image encoding. Step~4 follows these zeros to power (\figref{fig:flip-chain}).

The footprint follows the activations (Step~1), and after training the activations follow the content of the image. The footprint should carry a projection of the learned representation, and the target we regress to should itself be a learned representation, in which similar images lie close together, and not pixels or labels. This is the second design choice. These ReLU tests concern CNNs and ResNet-18; our CLIP ViT-L/14 victim uses QuickGELU~\cite{hendrycks2016gelu,radford2021clip}.

\begin{figure}[!htbp]
\centering
\includegraphics[width=\columnwidth]{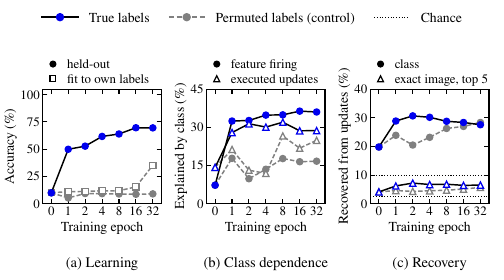}
\caption{Training raises recovery from executed operations.}
\Description{Three panels plot accuracy, the share of feature firing and executed updates explained by class, and class and exact-image recovery from executed updates over seven training checkpoints, for true and permuted training labels.}
\label{fig:learning-sweep}
\end{figure}

\paragraph{\normalfont\bfseries Step 3: Does Learning Make Executed Operations More Revealing?}
\label{root:learning-sweep}
If the footprint is a projection of the representation, what it reveals should grow as the representation is learned. We follow one four-layer CNN through 32 Imagenette epochs, once with true labels and once, as a control, with the labels permuted. At seven checkpoints, 1,000 held-out images run through a convolution kernel that skips zero activations, and the footprint is the kernel's exact count of executed vector updates: a count taken in software, not a PMU reading.
Learning raises what the executed-update count reveals about the input (\figref{fig:learning-sweep}). After one epoch, class explains 28.0\% of total executed updates, up from 14.3\%, and the class recovered from the update counts rises from 19.8\% to 28.9\% against 10\% chance. 
The permuted-label network stays at chance on held-out images.\looseness=-1

The work a zero-skipping kernel executes therefore follows what the network has learned, and a small network learns little. The target should thus be \textit{a representation learned at scale, and the attacker should not have to learn it. This is the third design choice: CHIP uses the frozen embedding of an encoder trained at scale, CLIP~\cite{radford2021clip}, and trains only the mapping from the footprint}.

\begin{figure}[!t]
\centering
\includegraphics[width=\columnwidth]{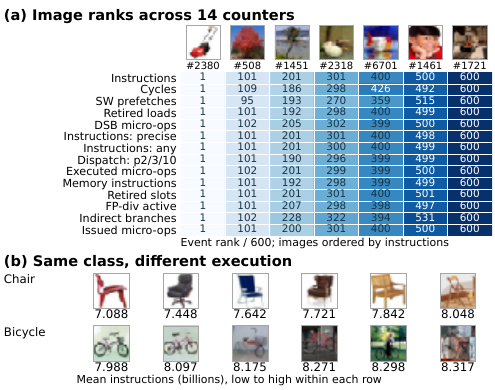}
\caption{Image and end-to-end execution variation.}
\Description{Seven images at evenly spaced instruction ranks are shown above their ranks across all fourteen supplied counters. A second panel shows all six chair images and all six bicycle images ordered by mean instructions.}
\label{fig:visual-footprint}
\end{figure}

\paragraph{\normalfont\bfseries Step 4: How Do Activation Values Reach Power, Time and Instruction Counts?}
\label{root:energy}
\label{root:flips}
Steps~1--3 concern what the activations contain; we now follow how they reach quantities an attacker can read. The first route is power. Activation values reach power through bit transitions~\cite{lipp2021platypus}: more zero activations cause fewer transitions and less energy (\figref{fig:flip-chain}). Total zero counts in an MNIST~\cite{lecun1998mnist} CNN identify digits at 25.3\% accuracy against 10\% chance. In a gate-level simulation of the same model synthesized with Yosys~\cite{wolfyosys}, digit 1 produces 4.4\% fewer transitions than digit 0, and ordering digits by bit transitions reproduces their classification-rank order exactly.
More aggressive microarchitectural optimization magnifies this effect: for example on Intel AMX~\cite{dizani2025thor}, 10 million 8-bit matrix multiplications consume 9.72~J with a zero operand versus 10.03--10.06~J with nonzero operands. Enabling or disabling Turbo Boost also changes many counter values, including prefetch counts.

\label{root:optim}
Power reaches time. Lower switching power can leave more headroom for Turbo Boost, changing elapsed time without changing retired instructions, as Hertzbleed~\cite{wang2022hertzbleed} demonstrates. On a pinned Xeon Gold 5420+ core, randomized Turbo Boost requests raise effective frequency by 0.25\% and lower elapsed time by 0.25\%, while instruction counts remain unchanged. The same comparison shows small mean shifts in \path{ref-cycles}, \path{uops_executed.core}, \path{uops_dispatched.port_0}, and \path{uops_dispatched.port_1}. Equal retired-instruction counts therefore do not imply equal timing or microarchitectural event counts.

The second route is the number of instructions the CPU executes and inference or generation latency itself. On \SPR{}, increasing the token-pruning threshold $\tau$ from 0.15 to 0.80 lowers median instructions from 8.58 to 4.74 billion and raises cross-image variation from 0.9\% to 4.6\%: pruning removes almost half of the instructions, and how much it removes depends on the image. CNN kernels that skip zero activations do the same~\cite{maji2024sparseleakynets}.

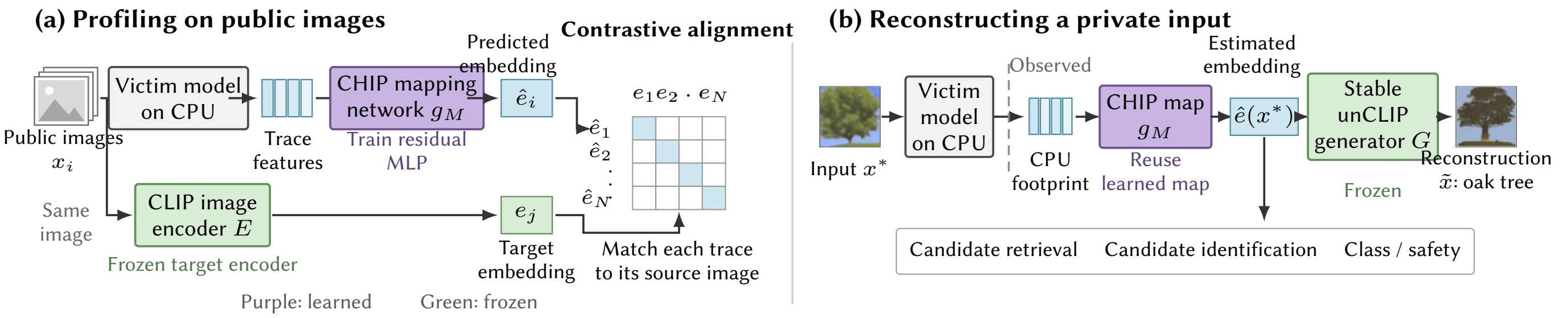
\begin{figure*}[!t]
\centering
\begingroup
\definecolor{chipgreen}{HTML}{D9ECD5}
\definecolor{chipgreenline}{HTML}{62925C}
\definecolor{chippurple}{HTML}{C9B5E3}
\definecolor{chippurpleline}{HTML}{6F4E91}
\definecolor{chipblue}{HTML}{CDE7F0}
\definecolor{chipblueline}{HTML}{508CA5}
\definecolor{chipgray}{HTML}{F1F2F4}
\resizebox{\textwidth}{!}{%
\begin{tikzpicture}[x=1cm,y=1cm,>=latex,
  font=\sffamily\fontsize{7}{8}\selectfont,
  chiparrow/.style={->,draw=black!80,line width=0.65pt},
  chipsmall/.style={font=\sffamily\fontsize{6.5}{7.3}\selectfont,align=center},
  chipmodel/.style={draw=black!70,line width=0.65pt,rounded corners=1pt,
    align=center,inner sep=2pt,minimum height=0.67cm}]
\useasboundingbox (-0.25,-0.08) rectangle (16.55,3.30);
\node[anchor=west,font=\sffamily\bfseries\fontsize{8}{9}\selectfont] at (0,3.07)
  {(a) Profiling on public images};
\node[anchor=west,font=\sffamily\bfseries\fontsize{8}{9}\selectfont] at (8.55,3.07)
  {(b) Reconstructing a private input};
\draw[black!20,line width=0.55pt] (8.28,0.02) -- (8.28,2.84);

\foreach \dx/\dy in {0.10/0.10,0.05/0.05,0/0}{
  \draw[fill=white,draw=black!55,line width=0.45pt]
    ({0.12+\dx},{1.95+\dy}) rectangle ({0.70+\dx},{2.47+\dy});}
\fill[black!15] (0.16,1.99) rectangle (0.66,2.43);
\fill[white] (0.28,2.32) circle (0.045);
\fill[black!35] (0.16,1.99) -- (0.34,2.23) -- (0.49,2.06)
  -- (0.55,2.14) -- (0.66,1.99) -- cycle;
\node[chipsmall] at (0.43,1.68) {Public images\\$x_i$};
\draw[chiparrow] (0.83,2.24) -- (1.12,2.24);
\node[chipmodel,fill=chipgray,minimum width=1.05cm] at (1.67,2.24)
  {Victim model\\on CPU};
\draw[chiparrow] (2.21,2.24) -- (2.52,2.24);
\foreach \i/\col in {0/chipblue,1/chipblue!55,2/chipblue!85,3/chipblue!45}{
  \draw[fill=\col,draw=chipblueline,line width=0.4pt]
    ({2.57+0.14*\i},2.04) rectangle ({2.69+0.14*\i},2.44);}
\node[chipsmall] at (2.85,1.68) {Trace\\features};
\draw[chiparrow] (3.15,2.24) -- (3.48,2.24);
\node[chipmodel,fill=chippurple,draw=chippurpleline,
  minimum width=1.25cm] at (4.13,2.24) {CHIP mapping\\network $g_M$};
\node[chipsmall,text=chippurpleline!80!black] at (4.13,1.68)
  {Train residual\\MLP};
\draw[chiparrow] (4.78,2.24) -- (5.12,2.24);
\node[draw=chipblueline,fill=chipblue,minimum width=0.55cm,
  minimum height=0.40cm,inner sep=1pt] at (5.42,2.24) {$\hat e_i$};
\node[chipsmall] at (5.22,2.70) {Predicted\\embedding};

\draw[chiparrow] (0.90,2.24) -- (0.90,0.98) -- (1.18,0.98);
\node[chipsmall,text=black!60,anchor=west] at (0.05,0.88) {Same\\image};
\node[chipmodel,fill=chipgreen,draw=chipgreenline,
  minimum width=1.45cm] at (1.93,0.98) {CLIP image\\encoder $E$};
\node[chipsmall,text=chipgreenline!80!black] at (1.93,0.43)
  {Frozen target encoder};
\draw[chiparrow] (2.68,0.98) -- (5.11,0.98);
\node[draw=chipgreenline,fill=chipgreen,minimum width=0.55cm,
  minimum height=0.40cm,inner sep=1pt] at (5.42,0.98) {$e_j$};
\node[chipsmall] at (5.42,0.48) {Target\\embedding};

\node[chipsmall,font=\sffamily\bfseries\fontsize{7}{8}\selectfont]
  at (7.04,2.96) {Contrastive alignment};
\foreach \row in {0,1,2,3}{
  \foreach \col in {0,1,2,3}{
    \ifnum\row=\col
      \fill[chipblue] ({6.56+0.25*\col},{1.79-0.25*\row})
        rectangle ({6.81+0.25*\col},{2.04-0.25*\row});
    \else
      \fill[white] ({6.56+0.25*\col},{1.79-0.25*\row})
        rectangle ({6.81+0.25*\col},{2.04-0.25*\row});
    \fi
    \draw[black!40,line width=0.35pt]
      ({6.56+0.25*\col},{1.79-0.25*\row})
        rectangle ({6.81+0.25*\col},{2.04-0.25*\row});}}
\foreach \col/\idx in {0/1,1/2,3/N}{
  \node[chipsmall] at ({6.685+0.25*\col},2.27) {$e_{\idx}$};}
\node[chipsmall] at (7.185,2.27) {$\cdots$};
\foreach \row/\idx in {0/1,1/2,3/N}{
  \node[chipsmall,anchor=east] at (6.47,{1.915-0.25*\row})
    {$\hat e_{\idx}$};}
\node[chipsmall,anchor=east] at (6.47,1.415) {$\vdots$};
\draw[chiparrow] (5.73,2.24) -- (5.96,2.24) -- (5.96,1.91) -- (6.12,1.91);
\draw[chiparrow] (5.73,0.98) -- (6.03,0.98) -- (6.03,0.79)
  -- (7.06,0.79) -- (7.06,1.02);
\node[chipsmall] at (7.04,0.47) {Match each trace\\to its source image};
\node[chipsmall,text=black!65] at (3.94,0.03)
  {Purple: learned \qquad Green: frozen};

\begin{scope}[shift={(8.55,0)}]
\node[inner sep=0pt,draw=black!70,line width=0.55pt] at (0.35,2.04)
  {\chipoakoriginal{0.64cm}};
\node[chipsmall] at (0.35,1.47) {Input $x^*$};
\draw[chiparrow] (0.69,2.04) -- (0.94,2.04);
\node[chipmodel,fill=chipgray,minimum width=0.92cm] at (1.42,2.04)
  {Victim\\model\\on CPU};
\draw[chiparrow] (1.90,2.04) -- (2.23,2.04);
\draw[black!45,densely dashed,line width=0.6pt] (2.06,1.45) -- (2.06,2.68);
\foreach \i/\col in {0/chipblue,1/chipblue!55,2/chipblue!85,3/chipblue!45}{
  \draw[fill=\col,draw=chipblueline,line width=0.4pt]
    ({2.28+0.12*\i},1.85) rectangle ({2.38+0.12*\i},2.24);}
\node[chipsmall] at (2.51,1.43) {CPU\\footprint};
\node[chipsmall,text=black!60] at (2.51,2.60) {Observed};
\draw[chiparrow] (2.78,2.04) -- (3.05,2.04);
\node[chipmodel,fill=chippurple,draw=chippurpleline,
  minimum width=1.14cm] at (3.64,2.04) {CHIP map\\$g_M$};
\node[chipsmall,text=chippurpleline!80!black] at (3.64,1.43)
  {Reuse\\learned map};
\draw[chiparrow] (4.23,2.04) -- (4.48,2.04);
\node[draw=chipblueline,fill=chipblue,minimum width=0.60cm,
  minimum height=0.40cm,inner sep=1pt] at (4.81,2.04) {$\hat e(x^*)$};
\node[chipsmall] at (4.68,2.69) {Estimated\\embedding};
\draw[chiparrow] (5.14,2.04) -- (5.30,2.04);
\node[chipmodel,fill=chipgreen,draw=chipgreenline,
  minimum width=1.12cm] at (5.98,2.04) {Stable\\unCLIP\\generator $G$};
\node[chipsmall,text=chipgreenline!80!black] at (5.98,1.25) {Frozen};
\draw[chiparrow] (6.66,2.04) -- (6.84,2.04);
\node[inner sep=0pt,draw=black!70,line width=0.55pt] at (7.20,2.04)
  {\chipoakdecoded{0.64cm}};
\node[chipsmall] at (7.20,1.47) {Reconstruction\\$\tilde x$: oak tree};
\draw[chiparrow] (4.81,1.82) -- (4.81,0.91);
\node[draw=black!35,fill=white,rounded corners=1pt,
  inner sep=4pt,chipsmall] at (3.96,0.59)
  {Candidate retrieval \quad Candidate identification \quad Class / safety};
\end{scope}
\end{tikzpicture}%
}
\endgroup
\caption{CHIP: \textbf{(a)} profiling learns the trace-to-embedding map; \textbf{(b)} the attack-time estimate conditions frozen unCLIP.}
\Description{Side-by-side panels show profiling and reconstruction. The training pipeline pairs public-image CPU traces with frozen CLIP embeddings to learn a residual projector through contrastive alignment. At attack time the projector estimates an embedding from a private trace; frozen unCLIP generates an image, and similarity supports retrieval, identification, and classification.}
\label{fig:generalized-chip-pipeline}
\end{figure*}

\paragraph{\normalfont\bfseries Step 5: Which Measurements Carry the Signal?}
\label{root:structure}
\label{root:association}
\label{paper:sec:event-statistics}
\label{paper:sec:counterdist}
The routes of Step~4 do not end in one counter. Instruction counts, data reuse, switching activity, frequency and token pruning each affect several pipeline stages, so all evaluated counters observe the same content-dependent computation. In a 500-image study, fourteen events vary together by 2.4--3.0\% across images. \figref[(a)]{fig:visual-footprint} ranks a 600-image set by instructions: sampled images hold similar, not identical, ranks across all fourteen events. In \figref[(b)]{fig:visual-footprint}, chairs span 7.088--8.048 billion instructions and bicycles 7.988--8.317 billion; the signature is not simply a class label, nor does one event uniquely identify a visual feature.

This is the fifth design choice: CHIP maps the joint signature across events and views to one whole-image embedding, and any subset that an interface exposes (time alone, or the cycle counters that survive enclave entry) uses the same mapping. \S\ref{paper:sec:measurements} measures how much each subset retains.
Learned features shape execution, and instruction-counting events and power-sensitive timing observe it in complementary ways. CHIP works with timing alone and with unprivileged performance counters on Intel processors; inside an Intel SGX enclave, the fixed cycle counters and elapsed time remain observable. Counter restrictions therefore still leave timing leakage visible. We report timing-only and unprivileged counter-based results in \S\ref{sec:paper-results}.

\subsection{Attack Building Blocks}
\label{sec:attack-building}

\figref{fig:generalized-chip-pipeline} shows the pipeline. CHIP separates estimating an embedding from synthesizing an image. Only one component is trained: a small residual multilayer perceptron, the mapping network, which converts CPU measurements into a CLIP embedding (panel~(a)). The CLIP image encoder supplying training targets and the Stable unCLIP generator producing images (panel~(b)) remain as released, without retraining. The same network architecture and target space serve every victim study; only the mapping is fitted to each measurement setting. The three building blocks follow their order of use.

\subsubsection{Estimating a Frozen CLIP Embedding}
\label{sec:attack-blocks}

\textbf{What the attacker records.} The victim processes each image as ten separate encoder calls, or views: the whole image and nine overlapping $224\times224$ crops on a $3\times3$ grid. For each view, the attacker records each selected hardware-event total, such as retired instructions or cycles, over the whole encoder call, repeats the measurement several times and averages the repetitions. Each image's trace is a list of numbers, ten views times the number of events (80 values on Haswell and 140 on Sapphire Rapids), kept in a fixed order. Each number is a total for one encoder call, not a layer-by-layer time series. The timing-only and restricted-counter variants record fewer or different numbers but keep the same target.\looseness=-1

\textbf{Locating the timed call.} The timed operation is the victim's image-encoder call, not the attacker's unCLIP generator. Timing requires observable invocation and completion boundaries: Akinsanya and Brennan~\cite{akinsanya2024timing} observe query/response timing, while Whispering MLaaS~\cite{shukla2023whispering} describes monitoring victim-process entry and exit by PID. The latter delimits a process, not an internal call in a persistent server. We require boundary visibility for each measured encoder call. Our evaluation uses controlled call boundaries: the collection harness releases one call after warm-up and receives its completion signal. Synchronization supplies the measured interval; running on another hardware thread alone does not. The experiment evaluates leakage given these boundaries, not discovery of hidden view boundaries.

\textbf{From durations to the input matrix.} Each interval contributes a duration: completion timestamp minus start timestamp. We average repeated observations of the same view and timing quantity, then concatenate the measurements in a fixed view/quantity order into one feature vector per image. One duration per view gives ten entries; views and repetitions are distinct. Stacking vectors from $N$ public images gives an $N\times d$ input matrix, with images as rows and measured features as columns, not pixels, layers or successive time samples. The internal-time ablation uses $d=20$ features; the feature count depends on the timing configuration, not on reshaping a single elapsed time. A test image supplies one $d$-entry row to the learned mapping, which estimates its 768-entry CLIP embedding. In the offline-profiling variant, the timing row first passes through linear calibration to estimated token counts.

\textbf{What the attacker predicts.} Each training image's target is its CLIP embedding: the 768 numbers that the frozen CLIP ViT-L/14 encoder~\cite{radford2021clip} assigns to the whole image, computed on the attacker's public images. Similar images have embeddings pointing in similar directions. The attacker never accesses the embedding of a private victim image.

\subsubsection{Training with Few Victim Traces}
\label{attack:training}
\textbf{Training on measured traces.} In the basic attack, the attacker runs the victim on its own public images, records each image's trace and computes its CLIP embedding. The mapping network learns from these pairs to estimate an embedding for a trace. Training combines four objectives. Two pull each estimate toward its own image's embedding: one reduces the squared difference between them (weight 0.1), and the other reduces their angle. The third is a contrastive objective, a symmetric adaptation of InfoNCE~\cite{oord2018cpc} with a learned temperature as in CLIP~\cite{radford2021clip}: within each training batch, every estimate must be closer to its own image's embedding than to the embeddings of the other images, and every embedding closer to its own estimate. The fourth penalizes estimates that collapse toward one average embedding (weight 0.5).\looseness=-1

\textbf{Public-image profiling and calibration.} 
Measuring the victim on many images is unrealistic, so CHIP's offline-profiling variant trains the mapping network entirely without the target victim CPU. 
The attacker runs a public copy of the ViT on its own machine and records how many tokens each of the three pruning stages processes in each view; these counts are exact and contain no hardware noise. Summing the three stages gives one count per view, or ten numbers per image, and the mapping network learns from these counts and the CLIP embeddings of public images. 
To apply the network on the target CPU, the attacker fits a linear regression converting target-CPU measurements into estimated token counts. At attack time, a measured trace passes through this conversion and the mapping network. Measured target-CPU traces are needed only to fit the conversion and select the training checkpoint on validation images, for each processor and event subset.

\subsubsection{Using the Estimated Embedding}
\label{attack:readouts}
\label{meth:pretrained}
\label{meth:metrics}
In the attack phase (\figref{fig:generalized-chip-pipeline}, panel~(b)), the attacker records a trace while the victim processes a private image and passes it through the trained mapping network, after the linear conversion in the offline variant. The estimate has four uses. For retrieval, the attacker compares the estimate with the CLIP embedding of each image in a candidate set by cosine similarity, which measures how closely two embeddings point in the same direction, and ranks the candidates. For identification, the same similarity decides whether a trace and a candidate image match. For classification, the attacker averages the training embeddings of each class into a prototype and assigns the class whose prototype is most similar; the same procedure gives fine-class, superclass and safe-versus-unsafe predictions without training another classifier. For reconstruction, frozen Stable unCLIP~\cite{stabilityunclip} generates an image directly from the estimate with a random seed, without substituting a retrieved candidate or accessing the private image. A separate frozen CLIP ViT-B/32 classifier evaluates the generated image.

\section{Methodology}
\label{sec:paper-methodology}
\noindent\textbf{Models and setup.}
\label{meth:setup}
CLIP ViT-L/14~\cite{radford2021clip} uses evaluation mode and eager attention, 256 patch tokens + [CLS] per $224\times224$ view; pruning~\cite{rao2021dynamicvit,yin2022avit} follows blocks 7/13/19 (zero-based 6/12/18). The CNN is ResNet-18~\cite{he2016resnet}. Stable unCLIP's encoder~\cite{stabilityunclip} embeds resized, normalized $224\times224$ whole images; ViT-B/32 labels generated images.\looseness=-1

\noindent\textbf{Images and partitions.}
\label{meth:data}
Official CIFAR-100~\cite{krizhevsky2009cifar} test images: Has\-well 1,500 (15/class), Sapphire Rapids 3,000 (30/class). Seed-zero within-class shuffling assigns five folds round-robin: each tests once, the next validates, three train (train, validation, test: 900/300/300 and 1,800/600/600). Test folds supply retrieval candidates. Single splits are balanced 60\%/20\%/20\%. Unsafe-Diffusion~\cite{qu2023unsafe} has 200 safe and 200 unsafe images (240/80/80); a Sapphire Rapids event study has 500 (350/75/75). Measurements are grouped by image.{\emergencystretch=1.5em\par}

The offline Haswell profile has 59,400 public pairs of token counts and embeddings without victim traces: 8,500 unmeasured CIFAR-100 test images, 50,000 official training images and 900 measured training images. It excludes fold validation/test images; ``held-out'' refers to attack, not pretrained-model, training.

\noindent\textbf{Measurement.}
\label{meth:traces}
A pinned thread runs ten encoder calls: one whole image and nine overlapping $224\times224$ crops on a $3\times3$ grid. Haswell (Xeon E5-2680 v3) records 8 events (2 groups, 5 repetitions, up to 12 concurrent victims); Sapphire Rapids (Xeon Gold 5420+) records 14 (3 groups, 10 repetitions, up to 26 victims).\looseness=-1

\noindent\textbf{Timing measurements (CHIP-T).}
The CPU harness times visible encoder-call boundaries internally; an unprivileged co-located process measures external elapsed time.

\noindent\textbf{Per-task counter measurements (CHIP-C).}
The parent runs and monitors the victim via \path{perf_event_open}, subject to task-access checks. Haswell uses \path{perf_event_paranoid}=2 for user-space per-task events; Sapphire Rapids uses level 1 for user/kernel events~\cite{linuxperfsecurity,perfeventopen}. FIFO-controlled \texttt{perf stat} starts after loading/warm-up, releases one call and stops on its completion signal. Validated groups give 80- or 140-entry mean traces (\S\ref{sec:attack-blocks}). Haswell ablations: ten instruction or twenty internal-time features, external co-located timing, or ordinary-process cycles.

\noindent\textbf{Enclave measurements (CHIP-E).}
SGX~\cite{intelsdm} protects memory from a compromised host OS. Opt-out suppresses retired-instruction and programmable counters; fixed cycles \path{IA32_FIXED_CTR1/2} remain active~\cite{intelsdm}. Our Gramine-SGX~\cite{tsai2017graphene} check finds nearly constant programmable counts and surviving cycles/time. TEEcorrelate~\cite{weissteiner2025teecorrelate} studies AMD SEV-SNP; PMU-Spill~\cite{qiu2023pmuspill} uses privileged OS access to SGX. Cycle/time retrieval uses non-enclave Haswell traces (\S\ref{root:structure}).\looseness=-1

\noindent\textbf{Event selection.}
\label{meth:counter-selection}
The PMU sweep excludes unsupported, zero-count, software and scheduler events. Sapphire Rapids validates 323 host-exposed core events; 202 in 70 groups yield 198 columns. Im\-age-grouped one-way ANOVA ranks between-image over repeated-run variation. We retain instructions/cycles and up to twelve programmable events (at most four programmable/two fixed per group; \texttt{perf stat} running $\geq99.5\%$). Sets: eight Haswell and fourteen Sapphire Rapids events; sweep and learning exclude evaluation.{\emergencystretch=1.5em\par}

\noindent\textbf{Training and selection.}
\label{meth:training}
Widths\slash residual blocks: 256/2 (five-fold\slash offline reconstruction), 1,024/3 (measured single-split), 128/2 (internal\slash external timing). AdamW~\cite{isca-v5:adamw} uses learning rate $3\times10^{-4}$, weight decay 0.01, batch $\leq256$, gradient clipping 1.0. Epoch limit\slash patience: 200/30 (validation top-five gain); single-split uses 300/40 with cosine scheduling. Training images fit linear calibration; validation selects regularization\slash checkpoints (\S\ref{attack:training}); test images are unused.{\emergencystretch=1.5em\looseness=-1\par}

Shuffled-target controls fix distributions, architecture, loss, budget, candidates and generator during retraining. Five-fold controls permute images with repetitions grouped, validation separately; single-split permutes trace records. Fixed test identities isolate pairing from marginals and generator prior. The reference feeds true, not estimated, CLIP embeddings to the same generator.\looseness=-1

\section{Evaluation}
\label{sec:paper-results}
This section summarizes our quantitative findings. 
\newcounter{resdetail}
\newcommand{\resultdetail}[1]{\stepcounter{resdetail}\paragraph*{\normalfont\itshape (\alph{resdetail})~#1}}

\definecolor{resattack}{HTML}{0078D4}
\definecolor{rescontrol}{HTML}{8B95A1}
\definecolor{resreference}{HTML}{111111}
\definecolor{resalternate}{HTML}{1F4E79}
\pgfplotsset{resaxis/.style={
 width=\columnwidth,height=2.52cm,
 axis lines=left,axis line style={black},tick style={black},
 ymin=0,enlarge x limits=0.22,
 tick align=outside,tick label style={font=\scriptsize,text=black},
 label style={font=\scriptsize,text=black},
 xticklabel style={align=center},
 ymajorgrids=true,grid style={black!12},
 legend style={font=\scriptsize,draw=black!35,fill=white,fill opacity=0.9,text opacity=1,
  at={(0.5,1.04)},anchor=south,legend columns=3,
  /tikz/every even column/.append style={column sep=4pt}},
 every axis plot/.append style={line width=0.7pt},
 every node near coord/.append style={font=\tiny,text=black},
 nodes near coords align={vertical},
 /pgf/number format/.cd,fixed,precision=3,
 /pgfplots/clip=false},
 resbar/.style={resaxis,ybar,bar width=7pt,nodes near coords},
 resxbar/.style={xbar,width=0.92\columnwidth,axis lines=left,axis line style={black},
  tick style={black},tick label style={font=\tiny,text=black},
  label style={font=\scriptsize,text=black},xmajorgrids=true,grid style={black!12},
  bar width=4.5pt,clip=false}}

\newsavebox{\rescropbox}
\newcommand{\rescrop}[6]{\begingroup
 \sbox{\rescropbox}{\includegraphics{#1}}%
 \edef\resleft{\the\dimexpr#2\wd\rescropbox\relax}%
 \edef\resbottom{\the\dimexpr#3\ht\rescropbox\relax}%
 \edef\resright{\the\dimexpr#4\wd\rescropbox\relax}%
 \edef\restop{\the\dimexpr#5\ht\rescropbox\relax}%
 \edef\rescommand{\noexpand\includegraphics[width=\the\dimexpr#6\relax,trim=\resleft\space\resbottom\space\resright\space\restop,clip]{#1}}%
 \rescommand\endgroup}

\setcounter{resdetail}{0}
\subsection{Can CPU Measurements Reconstruct the Image?}
\label{paper:sec:recon}
On held-out images, generated scenes recover object outlines, dominant colors and coarse layout; a classifier is correct at two to four times chance. For CLIP ViT-L/14, a residual map projects the CPU trace into frozen CLIP's 768-dimensional space; frozen Stable unCLIP synthesizes RGB images without candidates. 

\resultdetail{Held-out examples}
\figref[(a)]{paper:fig:examples} shows three Haswell CIFAR-100 images (300-image test split; offline mapping profiled on 59,400 public images). From trace-projected embeddings (\texttt{Estimated}), frozen unCLIP synthesizes a standing boy, vertical lamp post and upright bottle outline resembling the true-embedding reference; shuffled traces (\texttt{Shuffled}) yield unstructured texture.{\emergencystretch=1.5em\par}

\begin{figure}[!t]
\centering
\includegraphics[width=\columnwidth]{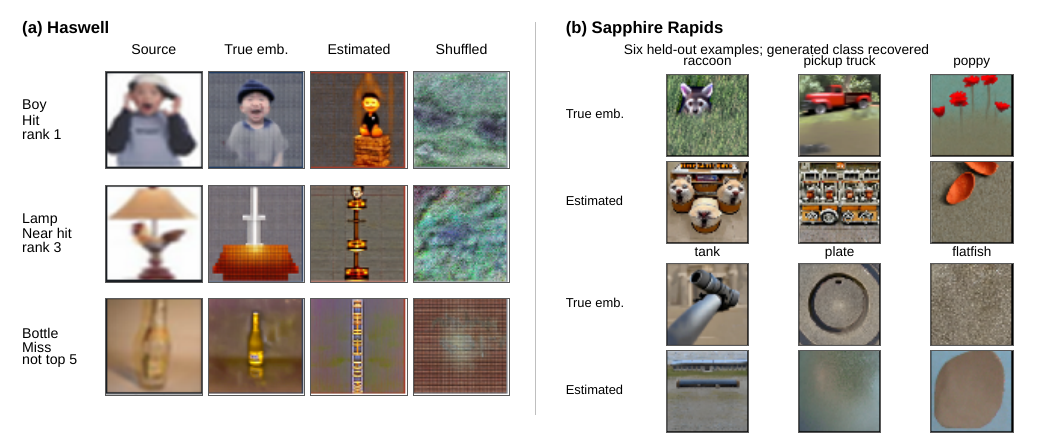}
\caption{Reconstructions: Haswell (a), Sapphire Rapids (b).}
\Description{Side-by-side processor panels use consistent labels and aligned images. Haswell retains all three source images, true-embedding generations, trace-estimated reconstructions and shuffled-trace controls, including hit, near-hit and miss ranks. Sapphire Rapids retains all six selected held-out examples whose generated class was recovered, each with its true-embedding generation and trace-estimated reconstruction.}
\label{paper:fig:examples}\label{paper:fig:examples-spr}
\end{figure}

\figref[(b)]{paper:fig:examples-spr} shows held-out Sapphire Rapids CIFAR-100 reconstructions without candidates, using the wider single-split map (\S\ref{meth:training}). Trace-projected embeddings (\texttt{Estimated}) generate a masked raccoon, multi-wheeled pickup, red poppy petals, angled tank barrel, circular plate and sandy flatfish shape. The top row gives true-embedding outputs. \figref{paper:fig:seen-path} additionally includes six training images, of which lookup exactly matches five; all other reconstruction and retrieval results use test images whose traces were excluded from mapping-network training.

\begin{figure}[!t]
\centering
\includegraphics[width=\columnwidth]{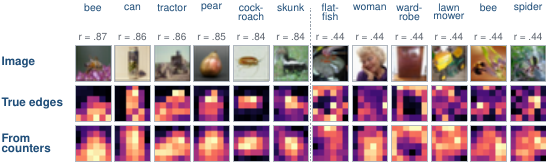}
\caption{CHIP's edge maps: true versus trace-estimated.}
\Description{Source edge maps and trace-estimated six-by-six edge maps, comparing coarse spatial structure.}
\label{paper:fig:spatial-edges}
\end{figure}

\resultdetail{Color, shape, and coarse spatial structure}
\label{paper:sec:color}
\label{paper:sec:shape}
Reconstructions recover most color available from the true embedding. CIFAR-100 per-channel RGB palette overlap rises from shuffled-trace $33.0\%$ to $43.8\%$ on Haswell ($1.33\times$), and from $30.3\%$ to $45.3\%$ on Sapphire Rapids ($1.50\times$), closing $75\%$ and $84\%$ of the gap to true-embedding generation. SSIM~\cite{wang2004ssim} is $0.04$ and $0.03$ on the two processors, versus shuffled-trace $0.02$ and $0.01$ and true-embedding $0.07\text{--}0.08$. Across processors, palette overlap gains $1.4\times$ and SSIM $2.5\times$. Absolute SSIM remains low even for true embeddings: unCLIP synthesizes a new image with similar content rather than inverting original pixels, so gains concern color and coarse structure. The twelve edge maps in \figref{paper:fig:spatial-edges} correlate with true maps at $r=0.44\text{--}0.87$.

\resultdetail{Classifying the reconstructed images}
\label{paper:sec:recon-classification}
A CLIP ViT-B/32 classifier measures generated-image class agreement within the CLIP family. Superclass accuracy is 11.7\% on Haswell and 10.8\% on Sapphire Rapids ($2.34\times$ and $2.16\times$ 5.0\% chance; shuffled 6.3\% and 4.2\%); fine-class accuracy is 4.0\% and 3.5\% ($4.0\times$ and $3.5\times$ 1.0\% chance; shuffled 1.7\% and 1.0\%).

Direct embedding classification (\S\ref{paper:sec:classification}), before image generation, separates the category signal from unCLIP's visual prior.

\label{paper:sec:seen-recon}

\setcounter{resdetail}{0}
\subsection{Can the Estimate Single Out the Exact Image?}
\label{paper:sec:retrieval}
Offline profiling ranks the exact source top-five for 24\% of Haswell queries ($14.5\times$ chance) and 56\% of Unsafe-Diffusion queries. Retrieval rises with profiling-set size and extends to another architecture.

\begin{figure}[!t]
\centering
\includegraphics[width=\columnwidth]{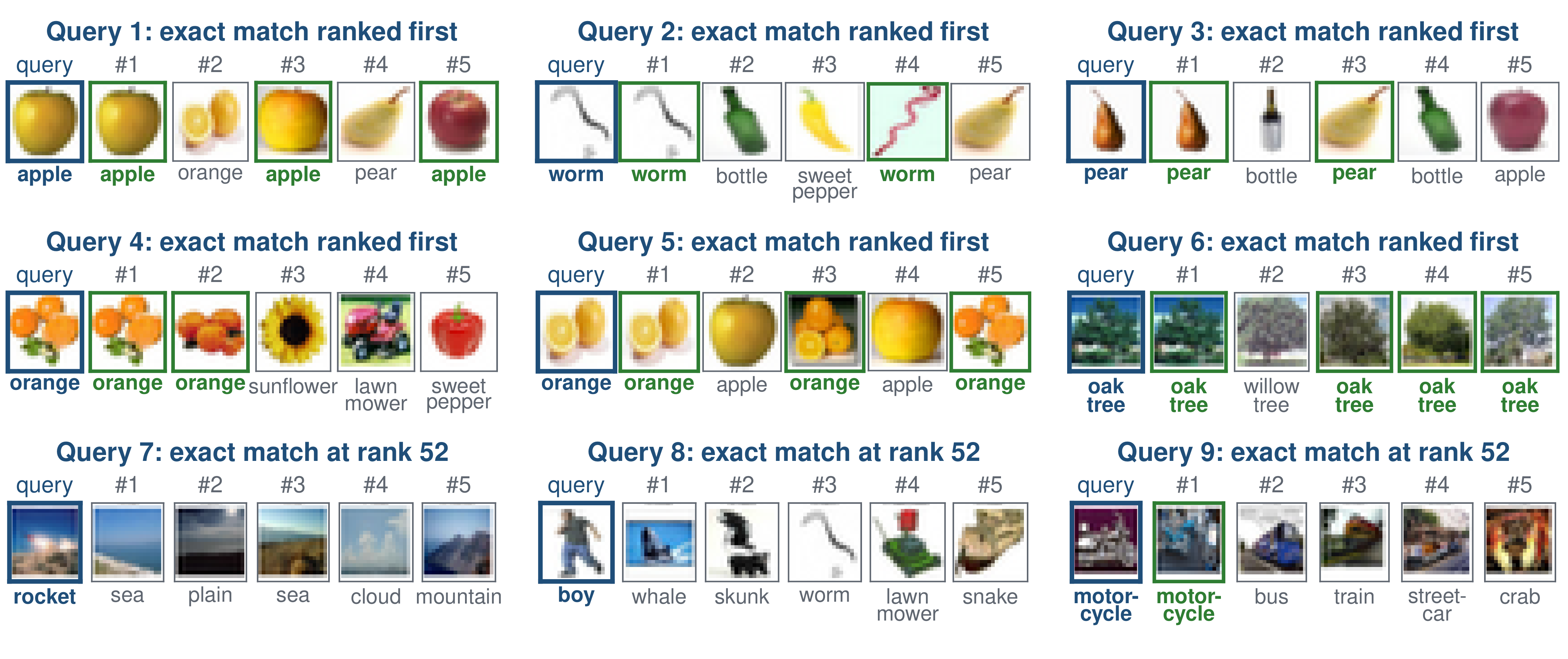}
\caption{CHIP's retrieval on ResNet-18 (2,000 candidates).}
\Description{Nine ResNet-18 queries and their retrieved candidates from a 2,000-image gallery; each panel reports the exact source rank.}
\label{paper:fig:resnet-retrieval}
\end{figure}

\resultdetail{Exact images in held-out candidate sets}
The basic attack, trained on measured traces (Table~\ref{paper:tab:retrieval-basic}), ranks the true image first for 3.47\% of Haswell queries ($10.4\times$ chance) and 2.40\% of Sapphire Rapids queries ($14.4\times$); top-five rates are 11.53\% ($6.9\times$) and 8.63\% ($10.4\times$). Unsafe-Diffusion reaches 12.0\% top-five versus 6.25\% chance ($1.92\times$; shuffled 5.0\%).

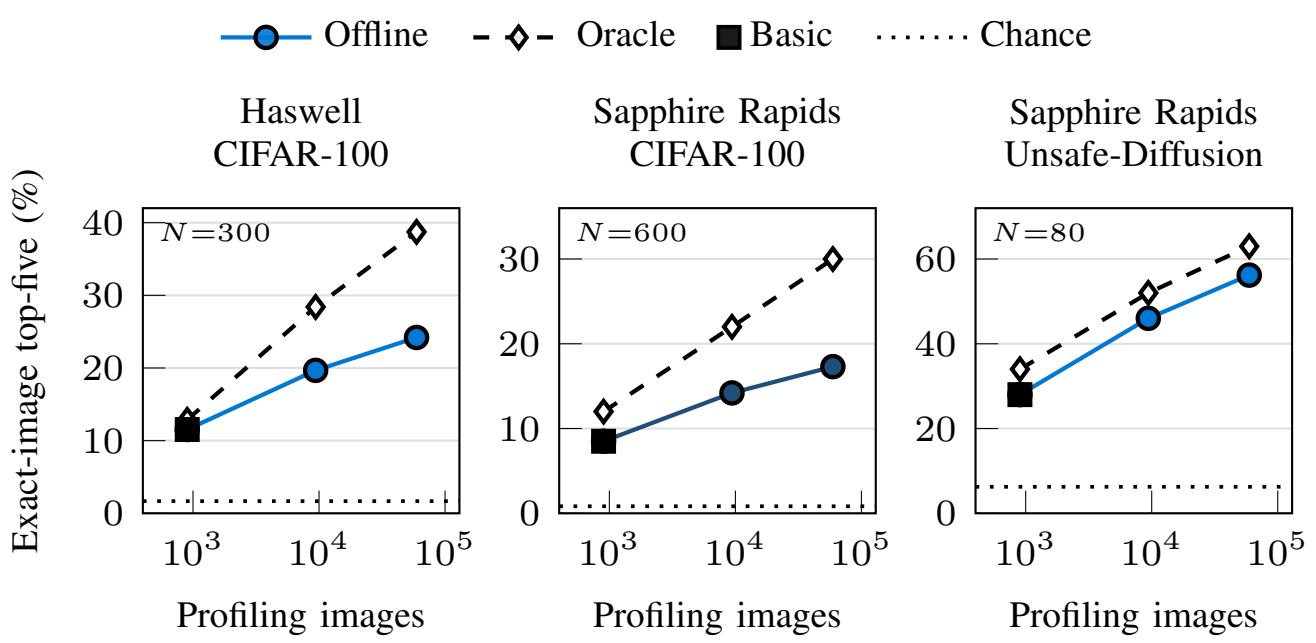
\begin{figure}[!t]
\centering
\pgfplotslegendfromname{paper:leg:prof}\\[1pt]
\begin{tikzpicture}
\begin{groupplot}[group style={group size=3 by 1,horizontal sep=0.66cm},
 resaxis,axis lines=box,xtick pos=bottom,ytick pos=left,tick align=inside,
 xlabel style={font=\scriptsize},ylabel style={font=\scriptsize},
 width=0.415\columnwidth,height=3.6cm,xmode=log,log basis x=10,
 xmin=400,xmax=130000,enlarge x limits=false,xtick={1000,10000,100000},
 xlabel={Profiling images},
 title style={font=\scriptsize,yshift=-4pt,align=center}, %
 legend style={at={(0.5,1.0)},anchor=south,legend columns=4,draw=none,fill=none,/tikz/every even column/.append style={column sep=6pt}}]
\nextgroupplot[title={Haswell\\CIFAR-100\strut},ymin=0,ymax=42,ylabel={Exact-image top-five (\%)},legend to name=paper:leg:prof]
\addplot[color=resattack,mark=*,mark options={draw=black,fill=resattack},thick] coordinates {(900,11.53) (9400,19.67) (59400,24.20)};
\addplot[color=black,dashed,mark=diamond*,mark options={solid,draw=black,fill=white},thick] coordinates {(900,12.87) (9400,28.40) (59400,38.73)};
\addplot[only marks,mark=square*,mark options={draw=black,fill=black}] coordinates {(900,11.53)};
\addplot[black,dotted,no markers] coordinates {(400,1.67) (130000,1.67)};
\node[anchor=north west,font=\tiny,inner sep=1.5pt] at (rel axis cs:0.02,0.98) {$N{=}300$};
\legend{Offline,Oracle,Basic,Chance}
\nextgroupplot[title={Sapphire Rapids\\CIFAR-100\strut},ymin=0,ymax=36]
\addplot[color=resalternate,mark=*,mark options={draw=black,fill=resalternate},thick] coordinates {(900,8.5) (9400,14.2) (59400,17.3)};
\addplot[color=black,dashed,mark=diamond*,mark options={solid,draw=black,fill=white},thick] coordinates {(900,12) (9400,22) (59400,30.0)};
\addplot[only marks,mark=square*,mark options={draw=black,fill=black}] coordinates {(900,8.5)};
\addplot[black,dotted,no markers] coordinates {(400,0.83) (130000,0.83)};
\node[anchor=north west,font=\tiny,inner sep=1.5pt] at (rel axis cs:0.02,0.98) {$N{=}600$};
\nextgroupplot[title={Sapphire Rapids\\Unsafe-Diffusion\strut},ymin=0,ymax=72]
\addplot[color=resattack,mark=*,mark options={draw=black,fill=resattack},thick] coordinates {(900,28) (9400,46) (59400,56.2)};
\addplot[color=black,dashed,mark=diamond*,mark options={solid,draw=black,fill=white},thick] coordinates {(900,34) (9400,52) (59400,63.0)};
\addplot[only marks,mark=square*,mark options={draw=black,fill=black}] coordinates {(900,28)};
\addplot[black,dotted,no markers] coordinates {(400,6.25) (130000,6.25)};
\node[anchor=north west,font=\tiny,inner sep=1.5pt] at (rel axis cs:0.02,0.98) {$N{=}80$};
\end{groupplot}
\end{tikzpicture}
\caption{CHIP's top-five retrieval under offline profiling.}
\Description{Three plots show exact-image top-five retrieval versus offline profiling-set size: Haswell on CIFAR-100, Sapphire Rapids on CIFAR-100, and Sapphire Rapids on Unsafe-Diffusion, with chance references.}
\label{paper:fig:profile-hsw-archive}
\label{paper:fig:profile-spr-archive}
\label{paper:fig:profile-nds-archive}
\end{figure}

\begin{table}[!t]
\centering\footnotesize
\setlength{\tabcolsep}{1.4pt}
\renewcommand{\arraystretch}{0.96}
\begin{tabular*}{\columnwidth}{@{\extracolsep{\fill}}lr*{3}{r}@{}}
\toprule
& & \multicolumn{2}{c}{\textbf{CIFAR-100}} & \shortstack{\textbf{Unsafe-}\\\textbf{Diffusion}} \\
\cmidrule(lr){3-4}\cmidrule(l){5-5}
\textbf{Metric} & \textbf{Top-\(k\)} & \textbf{Haswell} & \shortstack{\textbf{Sapphire}\\\textbf{Rapids}} & \shortstack{\textbf{Sapphire}\\\textbf{Rapids}} \\
& & \(N=300\) & \(N=600\) & \(N=80\) \\
\midrule
CHIP & 1 & 3.47 & 2.40 & 2.5 \\
& 5 & 11.53 & 8.63 & 12.0 \\
& 10 & 19.07 & 13.83 & 25.0 \\
\midrule
Shuffled & 1 & 0.80 & 0.10 & 1.5 \\
& 5 & 2.53 & 0.67 & 5.0 \\
& 10 & 4.00 & 1.63 & 11.0 \\
\midrule
Chance & 1 & 0.33 & 0.17 & 1.25 \\
& 5 & 1.67 & 0.83 & 6.25 \\
& 10 & 3.33 & 1.67 & 12.5 \\
\midrule
Gain (\%) & 1 & \(+941\) & \(+1340\) & \(+100\) \\
& 5 & \(+592\) & \(+936\) & \(+92\) \\
& 10 & \(+472\) & \(+730\) & \(+100\) \\
\bottomrule
\end{tabular*}

\caption{Basic retrieval (\%) and gain over chance.}
\label{paper:tab:retrieval-basic}
\end{table}
Shuffling trace--image pairs leaves retrieval near chance (2.53\% and 0.67\% top-five), attributing the gain to correct pairing. Ranking uses estimated embeddings directly, not the generator. On a common 75-image single-split candidate set, Haswell and Sapphire Rapids reach 29.9\% and 30.1\% top-five ($4.49\times$ and $4.52\times$ 6.67\% chance): processor generations eight years apart leak similarly.
\resultdetail{Offline profiling and finer-grained measurements}
\label{paper:sec:profiling}
\label{root:learning}
Offline profiling pairs token counts with embeddings from a public CLIP model on the attacker's machine; victim traces serve only calibration and checkpoint selection (\S\ref{meth:training}). On five-fold Haswell, adding 8,500 profiling images raises top-five retrieval from 11.53\% to 19.67\% ($11.8\times$ chance), and all 59,400 images give 24.20\% top-five and 8.40\% top-one ($14.5\times$ and $25.2\times$ chance), while the offline shuffled trace stays at 1.4\%--1.8\% (\figref{paper:fig:profile-hsw-archive}). Beyond 29,400 images the curve flattens at 23.2\%--24.5\%.

\label{paper:fig:saturation}

The flat region comes from aggregation, not network capacity. A four-times-wider mapping network at 59,400 images reaches 23.27\%, no gain over the default 24.20\%. Supplying the default network with all three per-stage counts per view instead raises top-five retrieval to 38.73\% ($23.2\times$) and top-one to 15.93\%. The thirty stage counts preserve at which pruning stage tokens are removed, which the ten per-view totals merge.

Offline top-one/top-five retrieval reaches $9.0\%/25.3\%$ on the single Haswell split (shuffled trace $2/300$ and $5/300$), $6.0\%/17.3\%$ on Sapphire Rapids CIFAR-100, and $31.2\%/56.2\%$ on Unsafe-Diffusion ($25.0\times/9.0\times$ chance; shuffled trace $2.5\%/8.8\%$). The per-stage count input reaches 30.0\% top-five on Sapphire Rapids and 63.0\% on Unsafe-Diffusion. Keeping only the sign of each of the 768 estimated coordinates gives 11.0\%, 9.2\%, and 52.5\% top-five on the three test sets ($6.6\times$, $11.0\times$, and $8.4\times$ chance), so the leaked embedding remains useful after one-bit quantization.

\label{paper:fig:profile-isca}

\resultdetail{Matched-input comparison with GAN}
\label{paper:sec:matched-gan}
With the same counter inputs, CHIP's frozen CLIP target retrieves the exact image about three times as often as a domain-trained GAN. We adapt the two-stage GAN of Yuan et al.~\cite{yuan2021private} to identical fourteen-counter inputs on a separate 600-image thumbnail set (300/100/200 training/validation/test images, three seeds). Here CHIP denotes the counter-to-CLIP retrieval stage, trained against frozen ViT-B/32 embeddings; GAN uses a counter encoder, a shared image decoder, and adversarial refinement at $32\times32$ resolution, a compact adaptation of the $128\times128$ access-trace attack. Both rank the same 200-image candidate set by frozen CLIP cosine, using CHIP's estimated embedding or the embedding of GAN's generated image, so the comparison measures the retrieval stage.

\begin{table}[!t]
\centering
\setlength{\tabcolsep}{3pt}
\begin{tabular*}{\columnwidth}{@{\extracolsep{\fill}}l*{5}{r}@{}}
\toprule
\textbf{Rank} & \textbf{CHIP} & \shortstack{\textbf{CHIP,}\\\textbf{matched}\\\textbf{capacity}} & \textbf{GAN} & \shortstack{\textbf{CHIP,}\\\textbf{shuffled}\\\textbf{pairs}} & \textbf{Chance} \\
\midrule
Top-1 & 2.17 & 1.83 & 0.67 & 0.50 & 0.50 \\
Top-5 & 6.50 & 6.67 & 2.33 & 2.17 & 2.50 \\
\bottomrule
\end{tabular*}
\caption{Matched-input retrieval, 200 candidates (\%).}
\label{paper:tab:matched-gan}
\end{table}

CHIP reaches 2.17\% top-one retrieval versus 0.67\% for GAN, and 6.50\% versus 2.33\% top-five (Table~\ref{paper:tab:matched-gan}). Reducing CHIP from 398,592 to 96,290 parameters, close to GAN's 95,631 total, retains 1.83\%/6.67\% top-one/top-five, so the difference does not come from model size.\looseness=-1

\resultdetail{Retrieval examples}
For three held-out Haswell queries, the boy ranks first, the lamp third, and the bottle outside the top five. Exact-image retrieval distinguishes photographs sharing the same class label.
A cross-core observer study on a 350/75/75 split gives 21/75 top-five hits (28.0\%, $4.2\times$ the 6.67\% chance; shuffled trace 5/75), and single, unaveraged traces give 27.6\%.

\resultdetail{Retrieval across victim architectures}
\label{root:architectures}
\label{paper:sec:resnet-retrieval}
ResNet-18~\cite{he2016resnet} has neither self-attention, token pruning, nor early exit, so it tests whether the leakage depends on pruning. In its 2,000-image candidate set, six of the nine queries shown rank the exact image first, and the other three rank it 52nd (\figref{paper:fig:resnet-retrieval}). Trace-to-embedding retrieval extends to a convolutional victim without pruning.

\setcounter{resdetail}{0}
\subsection{Can the Attacker Confirm That the Victim Processed an Image?}
\label{paper:sec:identify}
Identification compares matching and mismatched candidate--trace pairs, rather than ranking candidates (\S\ref{meth:metrics}). ROC AUC is near 0.85 on Intel processors and datasets: 0.851 on Haswell CIFAR-100, 0.848 on Sapphire Rapids CIFAR-100 and 0.857 on Unsafe-Diffusion, versus shuffled $0.494\text{--}0.501$ and 0.500 chance.

At 1\% FPR, CHIP identifies 21.0\%, 19.8\%, and 33.8\% of images ($21.0\times$, $19.8\times$, and $33.8\times$ chance); at 5\% FPR 42.0\%, 42.0\%, and 51.2\% (shuffled: Haswell 1.0\%/4.7\%, Unsafe-Diffusion 1.0\%/5.0\%).

AUC is similar on upscaled $32\times32$ CIFAR-100 inputs and on native higher-resolution Unsafe-Diffusion scenes, so the signal comes from image content not from CIFAR preprocessing.

\setcounter{resdetail}{0}
\subsection{Does the Estimate Reveal Class and Safety Labels?}
\label{paper:sec:classification}
The estimated embedding alone predicts the fine class at five to seven times chance and the safe/unsafe label at 66.2\% against 50\%, and its errors preserve the broad category.

\resultdetail{Classes predicted directly from the estimated embedding}
Class prototypes classify the estimated embedding, without generating an image (\S\ref{attack:readouts}). In the basic five-fold experiment, fine-class accuracy is 5.47\% on Haswell and 6.87\% on Sapphire Rapids ($5.5\times$ and $6.9\times$ the 1.0\% chance; shuffled trace 0.93\% and 0.77\%), and superclass accuracy is 13.93\% and 16.13\% ($2.8\times$ and $3.2\times$ the 5.0\% chance; shuffled trace 4.80\% and 4.53\%). Fine-class top-five accuracy reaches 18.33\% and 21.40\% against 5.0\% chance.

On one Sapphire Rapids split, when the fine class is wrong, 58 of 566 predictions fall in the correct superclass (10.2\% vs.\ $4/99=4.04\%$ chance), so the errors preserve broad category structure.

\resultdetail{Safe versus unsafe content}
\label{paper:sec:safety}
The safety attribute is a binary prototype decision on Unsafe-Diffusion. The split reaches 66.2\% accuracy against 50.0\% chance (shuffled trace 52.5\%), and the five-fold experiment reaches 60.2\% (shuffled trace 45.8\%).

\setcounter{resdetail}{0}
\subsection{Do Fewer Measurements or Counter Noise Stop the Attack?}
\label{subsec:robustness}
One measurement per image gives about 95\% of the ten-mea\-sure\-ment retrieval score; noise equal to the full between-image spread still leaves $6.6\times$ chance. We test two inexpensive defenses a provider could try: limiting how often a process observes the victim and perturbing its counters.{\emergencystretch=1.5em\par}

From one to ten measurements per image, top-ten retrieval rises from $12.86\%$ to $13.48\%$, top-five from $7.76\%$ to $8.08\%$, and top-one stays near $2.0\%$, about eight times chance (top-ten chance $1.67\%$) and eight to nine times the shuffled-trace control; the curve saturates by $N\!=\!2$. These runs omit single-crop augmentation.

Counter noise degrades the attack gradually. We inject zero-mean Gaussian noise of standard deviation $\alpha$ times the between-image spread, drawn independently for each read, so the attacker's repeat-averaging reduces it. Top-ten retrieval is $13.48\%$ without noise, $11.1\%$ at $\alpha\!=\!1$, where the noise equals the natural spread across images, and $5.42\%$ at $\alpha\!=\!8$ ($3.2\times$ chance), above the control throughout the sweep. Noise eight times the signal spread also removes the counters' monitoring value, so a lasting defense needs content-independent instruction counts and time (\S\ref{sec:mitigation}).

\resultdetail{Which measurements retain the signal?}
\label{paper:sec:measurements}
Removing measurements one at a time tests dependence on a single interface~\cite{cook2022biggerfish}; \S\ref{root:structure} gives reason to expect little loss.

Haswell basic retrieval reaches 11.47\% with instructions, 11.53\% with all counters, 10.87\% with internal time and 11.73\% with exact counts, against 1.67\% chance and 2.53\% for the shuffled trace with all counters; one whole-image trace gives 5.67\%. After a linear regression on instructions, other counters retain no above-chance residual, consistent with correlated measurements of one signal rather than independent channels.\looseness=-1

Offline retrieval reaches 12.1\% with cycles, 16.5\% with cycles plus time, 17.0\% with internal time and 16.3\% with external co-located timing, versus 24.2\% with all counters. Unsafe-Diffusion external timing reaches 15.7\% (6.25\% chance). The enclave check finds surviving cycles/time, but retrieval figures use ordinary inference traces; external timing needs no PMU.
\par

\section{Mitigations}
\label{sec:mitigation}
CHIP has no single interface to close. The image reaches the attacker through counters, through the cycle counters that survive enclave entry and through elapsed time (\S\ref{paper:sec:completeness}), so we discuss countermeasures at each layer where a processor vendor, operating system or inference provider can act. For each we state what it removes and what remains; Table~\ref{tab:mitigation} summarizes measured residuals.

\begin{table}[!t]
\centering\footnotesize\setlength{\tabcolsep}{3pt}
\renewcommand{\arraystretch}{1.1}
\begin{tabularx}{\columnwidth}{@{}>{\raggedright\arraybackslash}p{0.31\columnwidth}>{\raggedright\arraybackslash}p{0.19\columnwidth}>{\raggedright\arraybackslash}X@{}}
\toprule
\textbf{Mitigation} & \textbf{Removes} & \textbf{Remaining channel (Haswell, 1.67\% chance)} \\
\midrule
\path{perf_event_paranoid}=3 or no per-task PMU access & CHIP-C counters & Internal time 10.87\%; external time 16.3\% (offline) \\
\addlinespace[2pt]
Suppress fixed cycle counters on enclave entry & CHIP-E cycles & Ordinary inference timing proxy 17.0\% (offline) \\
\addlinespace[2pt]
One observed call per image & Ten-view sampling & One whole-image trace: 5.67\% \\
\addlinespace[2pt]
Bucketing per-crop token counts & Distinct token counts & 16-level bucketing: 9.9$\rightarrow$2.6 bits (ViT); 6.8$\rightarrow$2.3 bits/inf (Haswell); $\leq$1.7\% overhead \\
\bottomrule
\end{tabularx}
\caption{Counter restrictions and remaining channels.}
\label{tab:mitigation}
\end{table}

\vspace{2pt}\noindent{\textbf{Restricting access}}
After PLATYPUS~\cite{lipp2021platypus}, the Linux \texttt{powercap} driver was changed so that only root can read the RAPL energy counters~\cite{intel2020rapl}. The equivalent step is to raise \path{perf_event_paranoid} to 3 or to deny per-task counter access to tenants. This closes CHIP-C. It does not close CHIP-T: elapsed time needs no monitoring permission, and internal and external timing still retrieve the image (Table~\ref{tab:mitigation}). Unlike RAPL, the interface that remains is the clock, and it cannot be taken away from a client that waits for a response. Restriction also has a cost that RAPL did not have: profilers and counter-based attack detectors~\cite{ajorpaz2020perspectron,ajorpaz2022evax} read the same counters.

\vspace{2pt}\noindent{\textbf{Limiting resolution and adding noise}}
Reducing the granularity of the RAPL interface does not completely mitigate PLATYPUS~\cite{lipp2021platypus}, and the same holds. Counter noise eight times the between-image spread still leaves 5.42\% top-ten retrieval ($3.2\times$ chance, \S\ref{subsec:robustness}). Limiting the number of observations does not help either: one whole-image call still yields 5.67\% top-five ($3.4\times$ chance), and single cross-core traces reach 27.6\% against 28.0\% with averaging, so a defense must hold against individual traces. CHIP-T times an entire encoder call and not a single cache access, so we do not expect coarser timers to remove it; we did not evaluate them.

\vspace{2pt}\noindent{\textbf{Enclave counters}}
Opt-out enclave entry suppresses the retired-instruction and programmable counters but leaves fixed cycle counters active~\cite{intelsdm}, as our Gramine-SGX check confirms. For PLATYPUS, a microcode update changed the energy that RAPL reports when SGX is enabled~\cite{intel2020rapl}; a microcode update could likewise stop the fixed cycle counters on opt-out entry. As Lipp et al.~\cite{lipp2021platypus} note for RAPL, such a change needs trusted computing base recovery so that a remote verifier can check that it is in place. It would still leave CHIP-T: co-located timing matches in-victim timing (16.3\% vs 17.0\% top-five).

\vspace{2pt}\noindent{\textbf{Power management}}
One route from activation values to time passes through power and frequency (\S\ref{root:energy}). For Hertzbleed~\cite{wang2022hertzbleed}, Intel released no microcode update and instead issued software guidance for cryptographic code, such as masking and injected random delay~\cite{intel2022throttling}. Holding frequency fixed would remove this route at a performance cost but not the second: pruning and zero-skipping change the instruction count itself, and at a fixed input shape issued micro-ops change with content while retired micro-ops stay the same (\S\ref{root:noise}). We did not evaluate CHIP with Turbo Boost disabled.\looseness=-1

\vspace{2pt}\noindent{\textbf{Data-independent execution}}
For secrets, Intel's guidance is constant-time code, in which runtime, code access patterns and data access patterns are independent of the secret~\cite{intel2022consttime}, together with the data operand independent timing mode (DOITM)~\cite{intel2022doit}. In confidential inference the image is the secret, and token pruning, zero-skipping kernels and early exit make runtime depend on it by design. DOITM alone does not cover CHIP: its guidance states that power, frequency and performance-monitoring telemetry may still differ with data values processed in that mode~\cite{intel2022doit}.{\emergencystretch=1.5em\par}

\vspace{2pt}\noindent{\textbf{Token-count bucketing}}
Quantizing the surviving-token count of each crop into sixteen levels reduces the distinguishable execution patterns while retaining most of pruning's savings; bits denote the Shannon entropy of the realized per-image bucket assignment. On Sapphire Rapids, bucketing reduces the ViT token-count entropy from 9.94 to 2.59 bits over 60,000 images and the isolated pruning-region statistic from 9.49 to 3.58 bits over 1,520 images; on Haswell, it falls from 6.79 to 2.25 bits/inference. Overheads are 1.22\%, 1.22\% and 1.71\%, while 88.4\%, 94.8\% and 75.3\% of the pruning region's savings remain.

Five-fold linear retrieval on per-crop token counts gives 8.3--9.8\% versus 5.8--6.0\% shuffled control on 400 pruning-region images, and 1.6--1.9\% versus 0.7--0.9\% on 3,000 ViT images, across granularities including un-bucketed counts. Because time can vary while retired instructions stay fixed (Hertzbleed~\cite{wang2022hertzbleed}, \S\ref{root:noise}), constant instruction counts must be paired with constant time. Maya~\cite{pothukuchi2021maya} shapes power using control theory; CHIP calls for shaping instruction counts and time.

Following PLATYPUS and Hertzbleed precedents, a response could combine OS restrictions on tenant counter access (CHIP-C), microcode suppression of fixed cycle counters on opt-out enclave entry with trusted computing base recovery (CHIP-E), and framework guidance to treat inputs as secrets with constant-instruction-count, constant-time inference (CHIP-T). The first two are inexpensive but leave timing leakage; the third addresses content-dependent execution but loses some input-dependent optimization savings.

\section{Conclusion}
\label{sec:paper-conclusion}
CHIP maps CPU execution measurements into a frozen image embedding space. On the ViT victim, its estimates support scene synthesis without a candidate set, retrieval, identification, and classification on Haswell and Sapphire Rapids; a separate ResNet-18 study demonstrates exact-image retrieval. Retired instructions and elapsed time each retain retrieval information, so restricting a single counter interface does not remove the measured leakage. At a fixed input shape, image content changes issued micro-ops smoothly while retired instructions and micro-ops stay nearly constant, and training makes feature firing class selective; together these explain why execution leaks the image.\looseness=-1

\begin{acks}
Claude helped implement the GAN baseline of \S\ref{paper:sec:matched-gan}. Perplexity Computer helped restructure the manuscript, lay out the CPU timing and performance-counter analyses, and search the related literature. The authors reviewed all AI-assisted code and text and are responsible for the paper's experiments, results, citations and conclusions.
\end{acks}
\bibliographystyle{IEEEtran_doi}
\bibliography{refs}

@inproceedings{dosovitskiy2021vit,
  author    = {Dosovitskiy, Alexey and Beyer, Lucas and Kolesnikov, Alexander and
               Weissenborn, Dirk and Zhai, Xiaohua and Unterthiner, Thomas and
               Dehghani, Mostafa and Minderer, Matthias and Heigold, Georg and
               Gelly, Sylvain and Uszkoreit, Jakob and Houlsby, Neil},
  title     = {An Image is Worth 16x16 Words: Transformers for Image Recognition at Scale},
  booktitle = {Proc. Int. Conf. Learning Representations (ICLR)},
  year      = {2021}
}

@inproceedings{he2016resnet,
  author = {He, Kaiming and Zhang, Xiangyu and Ren, Shaoqing and Sun, Jian},
  title = {Deep Residual Learning for Image Recognition},
  booktitle = {Proc. IEEE Conf. Computer Vision and Pattern Recognition (CVPR)},
  pages = {770--778},
  year = {2016},
  doi = {10.1109/CVPR.2016.90}
}

@misc{stabilityunclip,
  author = {{Stability AI}},
  title = {Stable {unCLIP}},
  howpublished = {Stable Diffusion implementation documentation},
  year = {2023},
  url = {https://github.com/Stability-AI/stablediffusion/blob/main/doc/UNCLIP.MD},
  note = {Accessed: Nov.~13, 2026}
}

@article{wang2004ssim,
  author = {Wang, Zhou and Bovik, Alan Conrad and Sheikh, Hamid Rahim and Simoncelli, Eero P.},
  title = {Image Quality Assessment: From Error Visibility to Structural Similarity},
  journal = {IEEE Transactions on Image Processing},
  volume = {13},
  number = {4},
  pages = {600--612},
  year = {2004},
  doi = {10.1109/TIP.2003.819861}
}

@inproceedings{radford2021clip,
  author    = {Radford, Alec and Kim, Jong Wook and Hallacy, Chris and
               Ramesh, Aditya and Goh, Gabriel and Agarwal, Sandhini and
               Sastry, Girish and Askell, Amanda and Mishkin, Pamela and
               Clark, Jack and Krueger, Gretchen and Sutskever, Ilya},
  title     = {Learning Transferable Visual Models From Natural Language Supervision},
  booktitle = {Proc. Int. Conf. Machine Learning (ICML)},
  pages     = {8748--8763},
  year      = {2021}
}

@misc{ramesh2022unclip,
  author    = {Ramesh, Aditya and Dhariwal, Prafulla and Nichol, Alex and
               Chu, Casey and Chen, Mark},
  title     = {Hierarchical Text-Conditional Image Generation with {CLIP} Latents},
  howpublished = {arXiv:2204.06125},
  year      = {2022}
}

@misc{oord2018cpc,
  author    = {van den Oord, Aaron and Li, Yazhe and Vinyals, Oriol},
  title     = {Representation Learning with Contrastive Predictive Coding},
  howpublished = {arXiv:1807.03748},
  year      = {2018}
}

@inproceedings{isca-v5:adamw,
  author    = {Loshchilov, Ilya and Hutter, Frank},
  title     = {Decoupled Weight Decay Regularization},
  booktitle = {Proc. Int. Conf. Learning Representations (ICLR)},
  year      = {2019}
}

@inproceedings{krizhevsky2012alexnet,
  author    = {Krizhevsky, Alex and Sutskever, Ilya and Hinton, Geoffrey E.},
  title     = {{ImageNet} Classification with Deep Convolutional Neural Networks},
  booktitle = {Advances in Neural Information Processing Systems (NeurIPS)},
  volume    = {25},
  pages     = {1097--1105},
  year      = {2012}
}

@techreport{krizhevsky2009cifar,
  author      = {Krizhevsky, Alex},
  title       = {Learning Multiple Layers of Features from Tiny Images},
  institution = {University of Toronto},
  year        = {2009}
}

@inproceedings{qu2023unsafe,
  author    = {Qu, Yiting and Shen, Xinyue and He, Xinlei and Backes, Michael and
               Zannettou, Savvas and Zhang, Yang},
  title     = {Unsafe Diffusion: On the Generation of Unsafe Images and Hateful
               Memes From Text-To-Image Models},
  booktitle = {Proc. ACM SIGSAC Conf. Computer and Communications Security (CCS)},
  year      = {2023}
}

@inproceedings{das2019sok,
  author    = {Das, Sanjeev and Werner, Jan and Antonakakis, Manos and
               Polychronakis, Michalis and Monrose, Fabian},
  title     = {{SoK}: The Challenges, Pitfalls, and Perils of Using Hardware
               Performance Counters for Security},
  booktitle = {Proc. IEEE Symp. Security and Privacy (S\&P)},
  year      = {2019}
}

@inproceedings{ajorpaz2020perspectron,
  author    = {Mirbagher-Ajorpaz, Samira and Pokam, Gilles and
               Mohammadian-Koruyeh, Esmaeil and Garza, Elba and
               Abu-Ghazaleh, Nael and Jim{\'e}nez, Daniel A.},
  title     = {{PerSpectron}: Detecting Invariant Footprints of Microarchitectural
               Attacks with Perceptron},
  booktitle = {Proc. IEEE/ACM Int. Symp. Microarchitecture (MICRO)},
  pages     = {1124--1137},
  year      = {2020}
}

@inproceedings{ajorpaz2022evax,
  author    = {{Mirbagher Ajorpaz}, Samira and Moghimi, Daniel and
               Collins, Jeffrey Neal and Pokam, Gilles and Abu-Ghazaleh, Nael and
               Tullsen, Dean},
  title     = {{EVAX}: Towards a Practical, Pro-active \& Adaptive Architecture for
               High Performance \& Security},
  booktitle = {Proc. IEEE/ACM Int. Symp. Microarchitecture (MICRO)},
  pages     = {1218--1236},
  year      = {2022}
}

@inproceedings{demme2013malware,
  author    = {Demme, John and Maycock, Matthew and Schmitz, Jared and
               Tang, Adrian and Waksman, Adam and Sethumadhavan, Simha and
               Stolfo, Salvatore},
  title     = {On the Feasibility of Online Malware Detection with Performance Counters},
  booktitle = {Proc. ACM/IEEE Int. Symp. Computer Architecture (ISCA)},
  year      = {2013}
}

@inproceedings{naghibijouybari2018rendered,
  author    = {Naghibijouybari, Hoda and Neupane, Ajaya and Qian, Zhiyun and
               Abu-Ghazaleh, Nael},
  title     = {Rendered Insecure: {GPU} Side Channel Attacks are Practical},
  booktitle = {Proc. ACM SIGSAC Conf. Computer and Communications Security (CCS)},
  year      = {2018}
}

@misc{nvidia2018renderedresponse,
  author    = {{NVIDIA}},
  title     = {Security Notice: Response to {Rendered Insecure}},
  howpublished = {Technical Notice 4738},
  month     = feb,
  year      = {2019}
}

@inproceedings{lipp2016armageddon,
  author    = {Lipp, Moritz and Gruss, Daniel and Spreitzer, Raphael and
               Maurice, Cl{\'e}mentine and Mangard, Stefan},
  title     = {{ARMageddon}: Cache Attacks on Mobile Devices},
  booktitle = {Proc. USENIX Security Symp.},
  pages     = {549--564},
  year      = {2016}
}

@inproceedings{weissteiner2025teecorrelate,
  author    = {Weissteiner, Hannes and Rauscher, Fabian and
               Schr{\"o}der, Robin Leander and Juffinger, Jonas and Gast, Stefan and
               Wichelmann, Jan and Eisenbarth, Thomas and Gruss, Daniel},
  title     = {{TEEcorrelate}: An Information-Preserving Defense against
               Performance-Counter Attacks on {TEEs}},
  booktitle = {Proc. USENIX Security Symp.},
  pages     = {2481--2498},
  year      = {2025}
}

@misc{wang2021leakedweb,
  author    = {Wang, Han and Sayadi, Hossein and Sasan, Avesta and
               Rafatirad, Setareh and Homayoun, Houman},
  title     = {{Leaked-Web}: Accurate and Efficient Machine Learning-Based Website
               Fingerprinting Attack through Hardware Performance Counters},
  howpublished = {arXiv:2110.01202},
  year      = {2021}
}

@inproceedings{shusterman2019robust,
  author    = {Shusterman, Anatoly and Kang, Lachlan and Haskal, Yarden and
               Meltser, Yosef and Mittal, Prateek and Oren, Yossi and Yarom, Yuval},
  title     = {Robust Website Fingerprinting Through the Cache Occupancy Channel},
  booktitle = {Proc. USENIX Security Symp.},
  year      = {2019}
}

@inproceedings{zhang2023headset,
  author    = {Zhang, Yicheng and Slocum, Carter and Chen, Jiasi and
               Abu-Ghazaleh, Nael},
  title     = {It's All in Your Head(set): Side-Channel Attacks on {AR/VR} Systems},
  booktitle = {Proc. USENIX Security Symp.},
  year      = {2023}
}

@inproceedings{son2026ovrwatcher,
  author    = {Son, Seonghun and Mukherjee, Chandrika and Aburas, Reham Mohamed and
               Gulmezoglu, Berk and Celik, Z. Berkay},
  title     = {Side-channel Inference of User Activities in {AR/VR} Using {GPU} Profiling},
  booktitle = {Proc. Network and Distributed System Security Symp. (NDSS)},
  year      = {2026}
}

@inproceedings{gulmezoglu2017perfweb,
  author    = {Gulmezoglu, Berk and Zankl, Andreas and Eisenbarth, Thomas and
               Sunar, Berk},
  title     = {{PerfWeb}: How to Violate Web Privacy with Hardware Performance Events},
  booktitle = {Proc. European Symp. Research in Computer Security (ESORICS)},
  pages     = {80--97},
  year      = {2017},
  doi       = {10.1007/978-3-319-66399-9_5}
}

@inproceedings{wang2024gpuzip,
  author    = {Wang, Yingchen and Paccagnella, Riccardo and Gang, Zhao and
               Vasquez, Willy R. and Kohlbrenner, David and Shacham, Hovav and
               Fletcher, Christopher W.},
  title     = {{GPU.zip}: On the Side-Channel Implications of Hardware-Based
               Graphical Data Compression},
  booktitle = {Proc. IEEE Symp. Security and Privacy (S\&P)},
  year      = {2024}
}

@inproceedings{yuan2021private,
  author    = {Yuan, Yuanyuan and Wang, Shuai and Zhang, Junping},
  title     = {Private Image Reconstruction from System Side Channels Using
               Generative Models},
  booktitle = {Proc. Int. Conf. Learning Representations (ICLR)},
  year      = {2021}
}

@inproceedings{yuan2022automated,
  author    = {Yuan, Yuanyuan and Pang, Qi and Wang, Shuai},
  title     = {Automated Side Channel Analysis of Media Software with Manifold Learning},
  booktitle = {Proc. USENIX Security Symp.},
  pages     = {4419--4436},
  year      = {2022}
}

@inproceedings{yuan2025ciphersteal,
  author    = {Yuan, Yuanyuan and Liu, Zhibo and Deng, Sen and Chen, Yanzuo and
               Wang, Shuai and Zhang, Yinqian and Su, Zhendong},
  title     = {{CipherSteal}: Stealing Input Data from {TEE}-Shielded Neural
               Networks with Ciphertext Side Channels},
  booktitle = {Proc. IEEE Symp. Security and Privacy (S\&P)},
  year      = {2025}
}

@inproceedings{zhang2024demystifying,
  author    = {Zhang, Zhiyuan and Lai, Zhenzhi and Parampalli, Udaya},
  title     = {{R+R}: Demystifying {ML}-Assisted Side-Channel Analysis
               Framework: A Case of Image Reconstruction},
  booktitle = {Proc. Annual Computer Security Applications Conf. (ACSAC)},
  year      = {2024}
}

@inproceedings{yan2020cachetelepathy,
  author    = {Yan, Mengjia and Fletcher, Christopher W. and Torrellas, Josep},
  title     = {Cache Telepathy: Leveraging Shared Resource Attacks to Learn {DNN}
               Architectures},
  booktitle = {Proc. USENIX Security Symp.},
  year      = {2020}
}

@inproceedings{hu2020deepsniffer,
  author    = {Hu, Xing and Liang, Ling and Li, Shuangchen and Deng, Lei and
               Zuo, Pengfei and Ji, Yu and Xie, Xinfeng and Ding, Yufei and
               Liu, Chang and Sherwood, Timothy and Xie, Yuan},
  title     = {{DeepSniffer}: A {DNN} Model Extraction Framework Based on Learning
               Architectural Hints},
  booktitle = {Proc. Int. Conf. Architectural Support for Programming Languages
               and Operating Systems (ASPLOS)},
  year      = {2020}
}

@inproceedings{batina2019csinn,
  author    = {Batina, Lejla and Bhasin, Shivam and Jap, Dirmanto and Picek, Stjepan},
  title     = {{CSI NN}: Reverse Engineering of Neural Network Architectures Through
               Electromagnetic Side Channel},
  booktitle = {Proc. USENIX Security Symp.},
  year      = {2019}
}

@inproceedings{maia2022hear,
  author    = {Maia, Henrique Teles and Xiao, Chang and Li, Dingzeyu and
               Grinspun, Eitan and Zheng, Changxi},
  title     = {Can One Hear the Shape of a Neural Network?: Snooping the {GPU} via
               Magnetic Side Channel},
  booktitle = {Proc. USENIX Security Symp.},
  year      = {2022}
}

@inproceedings{horvath2025barracuda,
  author    = {Horvath, Peter and Chmielewski, {\L}ukasz and Weissbart, L{\'e}o and
               Batina, Lejla and Yarom, Yuval},
  title     = {{BarraCUDA}: Edge {GPUs} do Leak {DNN} Weights},
  booktitle = {Proc. USENIX Security Symp.},
  pages     = {4017--4034},
  year      = {2025}
}

@inproceedings{isca-v5:hua2018,
  author    = {Hua, Weizhe and Zhang, Zhiru and Suh, G. Edward},
  title     = {Reverse Engineering Convolutional Neural Networks Through
               Side-channel Information Leaks},
  booktitle = {Proc. ACM/ESDA/IEEE Design Automation Conf. (DAC)},
  year      = {2018}
}

@inproceedings{alam2019secure,
  author    = {Alam, Manaar and Mukhopadhyay, Debdeep},
  title     = {How Secure are Deep Learning Algorithms from Side-Channel based
               Reverse Engineering?},
  booktitle = {Proc. ACM/IEEE Design Automation Conf. (DAC)},
  year      = {2019},
  doi       = {10.1145/3316781.3322465}
}

@inproceedings{wang2022stealthy,
  author    = {Wang, Han and Hafiz, Syed Mahbub and Patwari, Kartik and
               Chuah, Chen-Nee and Shafiq, Zubair and Homayoun, Houman},
  title     = {Stealthy Inference Attack on {DNN} via Cache-based Side-Channel Attacks},
  booktitle = {Proc. Design, Automation and Test in Europe Conf. (DATE)},
  year      = {2022}
}

@article{shukla2023whispering,
  author    = {Shukla, Shubhi and Alam, Manaar and Bhattacharya, Sarani and
               Mitra, Pabitra and Mukhopadhyay, Debdeep},
  title     = {``{Whispering MLaaS}'': Exploiting Timing Channels to Compromise
               User Privacy in Deep Neural Networks},
  journal   = {IACR Trans. Cryptographic Hardware and Embedded Systems (TCHES)},
  volume    = {2023},
  number    = {2},
  pages     = {587--613},
  year      = {2023},
  doi       = {10.46586/tches.v2023.i2.587-613}
}

@article{maji2024sparseleakynets,
  author    = {Maji, Saurav and Lee, Kyungmi and Chandrakasan, Anantha P.},
  title     = {{SparseLeakyNets}: Classification Prediction Attack Over
               Sparsity-Aware Embedded Neural Networks Using Timing Side-Channel
               Information},
  journal   = {IEEE Computer Architecture Letters},
  volume    = {23},
  number    = {1},
  pages     = {133--136},
  year      = {2024}
}

@article{dizani2025thor,
  author    = {Dizani, Farshad and Ghanbari, Azam and Kalyanapu, Joshua and
               Asher, Darsh and {Mirbagher Ajorpaz}, Samira},
  title     = {Thor: A Non-Speculative Value Dependent Timing Side Channel Attack
               Exploiting {Intel AMX}},
  journal   = {IEEE Computer Architecture Letters},
  volume    = {24},
  number    = {1},
  pages     = {69--72},
  year      = {2025},
  doi       = {10.1109/LCA.2025.3544989}
}

@inproceedings{isca-v5:gatebleed,
  author    = {Kalyanapu, Joshua and Dizani, Farshad and Asher, Darsh and
               Ghanbari, Azam and Cammarota, Rosario and Aysu, Aydin and
               {Mirbagher Ajorpaz}, Samira},
  title     = {{GATEBLEED}: Exploiting On-Core Accelerator Power Gating for High
               Performance and Stealthy Attacks on {AI}},
  booktitle = {Proc. IEEE/ACM Int. Symp. Microarchitecture (MICRO)},
  pages     = {308--325},
  year      = {2025}
}

@article{asher2026featurebleed,
  author    = {Asher, Darsh and Dizani, Farshad and Kalyanapu, Joshua and
               Cammarota, Rosario and Aysu, Aydin and {Mirbagher Ajorpaz}, Samira},
  title     = {{FeatureBleed}: Inferring Private Enriched Attributes From
               Sparsity-Optimized {AI} Accelerators},
  journal   = {IEEE Computer Architecture Letters},
  volume    = {25},
  number    = {1},
  pages     = {77--80},
  year      = {2026}
}

@inproceedings{li2022multiexit,
  author    = {Li, Zheng and Liu, Yiyong and He, Xinlei and Yu, Ning and
               Backes, Michael and Zhang, Yang},
  title     = {Auditing Membership Leakages of Multi-Exit Networks},
  booktitle = {Proc. ACM SIGSAC Conf. Computer and Communications Security (CCS)},
  year      = {2022}
}

@inproceedings{akinsanya2024timing,
  author    = {Akinsanya, Ayomide and Brennan, Tegan},
  title     = {Timing Channels in Adaptive Neural Networks},
  booktitle = {Proc. Network and Distributed System Security Symp. (NDSS)},
  year      = {2024}
}

@inproceedings{wei2018iknow,
  author    = {Wei, Lingxiao and Luo, Bo and Li, Yu and Liu, Yannan and Xu, Qiang},
  title     = {I Know What You See: Power Side-Channel Attack on Convolutional
               Neural Network Accelerators},
  booktitle = {Proc. Annual Computer Security Applications Conf. (ACSAC)},
  year      = {2018}
}

@article{moini2021bnn,
  author    = {Moini, Shayan and Tian, Shanquan and Holcomb, Daniel and
               Szefer, Jakub and Tessier, Russell},
  title     = {Power Side-Channel Attacks on {BNN} Accelerators in Remote {FPGAs}},
  journal   = {IEEE J. Emerging and Selected Topics in Circuits and Systems},
  volume    = {11},
  number    = {2},
  pages     = {357--370},
  year      = {2021}
}

@article{maji2021leakynets,
  author    = {Maji, Saurav and Banerjee, Utsav and Chandrakasan, Anantha P.},
  title     = {Leaky Nets: Recovering Embedded Neural Network Models and Inputs
               through Simple Power and Timing Side-Channels -- Attacks and Defenses},
  journal   = {IEEE Internet of Things Journal},
  volume    = {8},
  number    = {15},
  pages     = {12079--12092},
  year      = {2021},
  doi       = {10.1109/JIOT.2021.3061314}
}

@inproceedings{gao2025iknow,
  author    = {Gao, Zibo and Hu, Junjie and Guo, Feng and Zhang, Yixin and
               Han, Yinglong and Liu, Siyuan and Li, Haiyang and Lv, Zhiqiang},
  title     = {I Know What You Said: Unveiling Hardware Cache Side-Channels in
               Local Large Language Model Inference},
  booktitle = {Proc. USENIX Security Symp.},
  pages     = {1649--1668},
  year      = {2025}
}

@inproceedings{qiu2026keytar,
  author    = {Qiu, Mufan and Chuang, Lihsuan and Kim, Dohhyun and Qu, Huaizhi and
               Chen, Tianlong and Kwong, Andrew},
  title     = {{KeyTAR}: Practical Keystroke Timing Attacks and Input Reconstruction},
  booktitle = {Proc. IEEE Symp. Security and Privacy (S\&P)},
  year      = {2026},
  doi       = {10.1109/SP63933.2026.00106}
}

@misc{carlini2024remote,
  author    = {Carlini, Nicholas and Nasr, Milad},
  title     = {Remote Timing Attacks on Efficient Language Model Inference},
  howpublished = {arXiv:2410.17175},
  year      = {2024}
}

@misc{zhang2024timewilltell,
  author    = {Zhang, Tianchen and Saileshwar, Gururaj and Lie, David},
  title     = {Time Will Tell: Timing Side Channels via Output Token Count in
               Large Language Models},
  howpublished = {arXiv:2412.15431},
  year      = {2024}
}

@inproceedings{hadad2026shape,
  author    = {Hadad, Eyal and Guri, Mordechai},
  title     = {Shape and Substance: Dual-Layer Side-Channel Attacks on Local
               Vision-Language Models},
  booktitle = {arXiv preprint arXiv:2603.25403},
  year      = {2026}
}

@inproceedings{ding2025moecho,
  author    = {Ding, Ruyi and Xu, Tianhong and Shen, Xinyi and
               Ding, Aidong Adam and Fei, Yunsi},
  title     = {{MoEcho}: Exploiting Side-Channel Attacks to Compromise User
               Privacy in Mixture-of-Experts {LLMs}},
  booktitle = {Proc. ACM SIGSAC Conf. Computer and Communications Security (CCS)},
  pages     = {2159--2173},
  year      = {2025},
  doi       = {10.1145/3719027.3765174}
}

@misc{linuxperfsecurity,
  author    = {{The Linux Kernel Development Community}},
  title     = {Perf Events and Tool Security},
  howpublished = {Linux kernel documentation},
  year      = {2026},
  url       = {https://www.kernel.org/doc/html/latest/admin-guide/perf-security.html},
  note      = {Accessed: Nov.~8, 2026}
}

@misc{perfeventopen,
  author    = {{Linux man-pages project}},
  title     = {perf\_event\_open(2)},
  howpublished = {Linux manual pages},
  year      = {2026},
  url       = {https://man7.org/linux/man-pages/man2/perf_event_open.2.html},
  note      = {Accessed: Nov.~8, 2026}
}

@misc{intelsdm,
  author    = {{Intel Corporation}},
  title     = {{Intel} 64 and {IA-32} Architectures Software Developer's Manual,
               Volume {3D}: System Programming Guide, Part 4},
  year      = {2026},
  url       = {https://cdrdv2-public.intel.com/671269/332831-sdm-vol-3d.pdf},
  note      = {Section 40.6.3, Performance Monitoring with Opt-out Entry.
               Accessed: Nov.~8, 2026}
}

@misc{micay2016perfparanoid,
  author    = {Micay, Daniel},
  title     = {Re: Re: [PATCH 1/2] security, perf: allow further restriction of perf\_event\_open},
  howpublished = {Linux kernel-hardening mailing list},
  month     = aug,
  year      = {2016},
  url       = {https://openwall.com/lists/kernel-hardening/2016/08/02/11},
  note      = {Discussion of the proposed perf\_event\_paranoid=3 restriction}
}

@article{qiu2023pmuspill,
  author    = {Qiu, Pengfei and Gao, Qiang and Liu, Chang and Wang, Dongsheng and
               Lyu, Yongqiang and Li, Xiaoyong and Wang, Chunlu and Qu, Gang},
  title     = {{PMU-Spill}: A New Side Channel for Transient Execution Attacks},
  journal   = {IEEE Transactions on Circuits and Systems I: Regular Papers},
  volume    = {70},
  number    = {12},
  pages     = {5048--5059},
  doi       = {10.1109/TCSI.2023.3298913},
  year      = {2023},
}

@inproceedings{cook2022biggerfish,
  author = {Cook, Jack and Drean, Jules and Behrens, Jonathan and Yan, Mengjia},
  title = {There's Always a Bigger Fish: A Clarifying Analysis of a Machine-Learning-Assisted Side-Channel Attack},
  booktitle = {Proc. Int. Symp. Computer Architecture (ISCA)},
  year = {2022},
  doi = {10.1145/3470496.3527416}
}

@inproceedings{lipp2021platypus,
  author = {Lipp, Moritz and Kogler, Andreas and Oswald, David and Schwarz, Michael and Easdon, Catherine and Canella, Claudio and Gruss, Daniel},
  title = {{PLATYPUS}: Software-based Power Side-Channel Attacks on x86},
  booktitle = {Proc. IEEE Symp. Security and Privacy (S\&P)},
  year = {2021}
}

@inproceedings{wang2022hertzbleed,
  author = {Wang, Yingchen and Paccagnella, Riccardo and He, Elizabeth Tang and Shacham, Hovav and Fletcher, Christopher W. and Kohlbrenner, David},
  title = {Hertzbleed: Turning Power Side-Channel Attacks Into Remote Timing Attacks on x86},
  booktitle = {Proc. USENIX Security Symp.},
  year = {2022}
}

@inproceedings{schwarz2019netspectre,
  author = {Schwarz, Michael and Schwarzl, Martin and Lipp, Moritz and Masters, Jon and Gruss, Daniel},
  title = {{NetSpectre}: Read Arbitrary Memory over Network},
  booktitle = {Proc. European Symp. Research in Computer Security (ESORICS)},
  pages = {279--299},
  year = {2019},
  doi = {10.1007/978-3-030-29959-0_14}
}

@inproceedings{taneja2023hotpixels,
  author = {Taneja, Hritvik and Kim, Jason and Xu, Jie Jeff and van Schaik, Stephan and Genkin, Daniel and Yarom, Yuval},
  title = {Hot Pixels: Frequency, Power, and Temperature Attacks on {GPUs} and {Arm} {SoCs}},
  booktitle = {Proc. USENIX Security Symp.},
  year = {2023}
}

@inproceedings{pothukuchi2021maya,
  author = {Pothukuchi, Raghavendra Pradyumna and Pothukuchi, Sweta Yamini and Voulgaris, Petros and Schwing, Alexander and Torrellas, Josep},
  title = {Maya: Using Formal Control to Obfuscate Power Side Channels},
  booktitle = {Proc. Int. Symp. Computer Architecture (ISCA)},
  year = {2021},
  doi = {10.1109/ISCA52012.2021.00074}
}

@inproceedings{deng2009imagenet,
  title={{ImageNet}: A Large-Scale Hierarchical Image Database},
  author={Deng, Jia and Dong, Wei and Socher, Richard and Li, Li-Jia and Li, Kai and Fei-Fei, Li},
  booktitle={Proc. IEEE Conf. Computer Vision and Pattern Recognition (CVPR)},
  pages={248--255},
  year={2009}
}

@incollection{hinton1986distributed,
 author={Hinton, Geoffrey E. and McClelland, James L. and Rumelhart, David E.},
 title={Distributed Representations},
 booktitle={Parallel Distributed Processing: Explorations in the Microstructure of Cognition, Volume 1: Foundations},
 editor={Rumelhart, David E. and McClelland, James L.},
 publisher={MIT Press},
 address={Cambridge, MA},
 pages={77--109},
 year={1986}
}

@article{ackley1985boltzmann,
  author  = {Ackley, David H. and Hinton, Geoffrey E. and Sejnowski, Terrence J.},
  title   = {A Learning Algorithm for {Boltzmann} Machines},
  journal = {Cognitive Science},
  volume  = {9},
  number  = {1},
  pages   = {147--169},
  year    = {1985},
  doi     = {10.1207/s15516709cog0901_7}
}

@article{hinton2006deep,
 author={Hinton, Geoffrey E. and Osindero, Simon and Teh, Yee-Whye},
 title={A Fast Learning Algorithm for Deep Belief Nets},
 journal={Neural Computation}, volume={18}, number={7}, pages={1527--1554}, year={2006}
}

@inproceedings{hinton1993autoencoders,
 author={Hinton, Geoffrey E. and Zemel, Richard S.},
 title={Autoencoders, Minimum Description Length and {Helmholtz} Free Energy},
 booktitle={Advances in Neural Information Processing Systems (NeurIPS)}, volume={6}, pages={3--10}, year={1993}
}

@article{olshausen1996,
 author={Olshausen, Bruno A. and Field, David J.},
 title={Emergence of Simple-Cell Receptive Field Properties by Learning a Sparse Code for Natural Images},
 journal={Nature}, volume={381}, pages={607--609}, year={1996}
}

@incollection{barlow1961,
 author={Barlow, Horace B.},
 title={Possible Principles Underlying the Transformations of Sensory Messages},
 booktitle={Sensory Communication}, editor={Rosenblith, Walter A.}, publisher={MIT Press}, pages={217--234}, year={1961}
}

@article{barlow1972,
 author={Barlow, Horace B.},
 title={Single Units and Sensation: A Neuron Doctrine for Perceptual Psychology?},
 journal={Perception}, volume={1}, number={4}, pages={371--394}, year={1972}
}

@article{barlow2001,
 author={Barlow, Horace B.},
 title={Redundancy Reduction Revisited},
 journal={Network: Computation in Neural Systems}, volume={12}, number={3}, pages={241--253}, year={2001}
}

@article{rao1999,
 author={Rao, Rajesh P. N. and Ballard, Dana H.},
 title={Predictive Coding in the Visual Cortex: A Functional Interpretation of Some Extra-Classical Receptive-Field Effects},
 journal={Nature Neuroscience}, volume={2}, number={1}, pages={79--87}, year={1999}
}

@article{friston2007,
 author={Friston, Karl J. and Stephan, Klaas E.},
 title={Free-Energy and the Brain},
 journal={Synthese}, volume={159}, number={3}, pages={417--458}, year={2007}
}

@article{levy1996,
 author={Levy, William B. and Baxter, Robert A.},
 title={Energy Efficient Neural Codes},
 journal={Neural Computation}, volume={8}, number={3}, pages={531--543}, year={1996}
}

@inproceedings{hazelwood2018applied,
  author    = {Hazelwood, Kim and Bird, Sarah and Brooks, David and
               Chintala, Soumith and Diril, Utku and Dzhulgakov, Dmytro and
               Fawzy, Mohamed and Jia, Bill and Jia, Yangqing and
               Kalro, Aditya and Law, James and Lee, Kevin and Lu, Jason and
               Noordhuis, Pieter and Smelyanskiy, Misha and Xiong, Liang and
               Wang, Xiaodong},
  title     = {Applied Machine Learning at {Facebook}: A Datacenter
               Infrastructure Perspective},
  booktitle = {Proc. IEEE Int. Symp. High Performance Computer Architecture (HPCA)},
  pages     = {620--629},
  year      = {2018},
  doi       = {10.1109/HPCA.2018.00059}
}

@inproceedings{gao2024deeptheft,
  author    = {Gao, Yansong and Qiu, Huming and Zhang, Zhi and Wang, Binghui and
               Ma, Hua and Abuadbba, Alsharif and Xue, Minhui and Fu, Anmin and
               Nepal, Surya},
  title     = {{DeepTheft}: Stealing {DNN} Model Architectures through Power
               Side Channel},
  booktitle = {Proc. IEEE Symp. Security and Privacy (S\&P)},
  year      = {2024},
  doi       = {10.1109/SP54263.2024.00250}
}

@inproceedings{majumdar2022characterizing,
  author    = {Majumdar, Saikat and Samavatian, Mohammad Hossein and
               Teodorescu, Radu},
  title     = {Characterizing Side-Channel Leakage of {DNN} Classifiers
               though Performance Counters},
  booktitle = {Proc. IEEE Int. Symp. Hardware Oriented Security and Trust (HOST)},
  pages     = {45--48},
  year      = {2022},
  doi       = {10.1109/HOST54066.2022.9839882}
}

@inproceedings{srivastava2026ghosttrace,
  author    = {Srivastava, Amisha and Menon, Anand and Das, Sanjay and
               Basu, Kanad},
  title     = {{GhostTrace}: A Membership Inference Attack via Hardware
               Performance Counters as a Side-Channel},
  booktitle = {Proc. IEEE Int. Symp. Hardware Oriented Security and Trust (HOST)},
  pages     = {220--230},
  year      = {2026},
  doi       = {10.1109/HOST68814.2026.11604845}
}

@inproceedings{xu2015controlled,
  author    = {Xu, Yuanzhong and Cui, Weidong and Peinado, Marcus},
  title     = {Controlled-Channel Attacks: Deterministic Side Channels for
               Untrusted Operating Systems},
  booktitle = {Proc. IEEE Symp. Security and Privacy (S\&P)},
  pages     = {640--656},
  year      = {2015},
  doi       = {10.1109/SP.2015.45}
}

@inproceedings{gast2025counterseveillance,
  author    = {Gast, Stefan and Weissteiner, Hannes and
               Schr{\"o}der, Robin Leander and Gruss, Daniel},
  title     = {{CounterSEVeillance}: Performance-Counter Attacks on
               {AMD SEV-SNP}},
  booktitle = {Proc. Network and Distributed System Security Symp. (NDSS)},
  year      = {2025},
  doi       = {10.14722/ndss.2025.241038}
}

@inproceedings{scotti2023mindeye,
  author    = {Scotti, Paul S. and Banerjee, Atmadeep and Goode, Jimmie and
               Shabalin, Stepan and Nguyen, Alex and Cohen, Ethan and
               Dempster, Aidan J. and Verlinde, Nathalie and Yundler, Elad and
               Weisberg, David and Norman, Kenneth A. and Abraham, Tanishq Mathew},
  title     = {Reconstructing the Mind's Eye: {fMRI}-to-Image with Contrastive
               Learning and Diffusion Priors},
  booktitle = {Advances in Neural Information Processing Systems (NeurIPS)},
  year      = {2023}
}

@inproceedings{huegle2023power2picture,
  author    = {Huegle, Lukas and Gotthard, Martin and Meyers, Vincent and
               Krautter, Jonas and Gnad, Dennis R. E. and Tahoori, Mehdi B.},
  title     = {{Power2Picture}: Using Generative {CNNs} for Input Recovery of
               Neural Network Accelerators through Power Side-Channels on {FPGAs}},
  booktitle = {Proc. IEEE Int. Symp. Field-Programmable Custom Computing
               Machines (FCCM)},
  pages     = {155--161},
  year      = {2023},
  doi       = {10.1109/FCCM57271.2023.00025}
}

@inproceedings{rao2021dynamicvit,
  author    = {Rao, Yongming and Zhao, Wenliang and Liu, Benlin and Lu, Jiwen and
               Zhou, Jie and Hsieh, Cho-Jui},
  title     = {{DynamicViT}: Efficient Vision Transformers with Dynamic Token
               Sparsification},
  booktitle = {Advances in Neural Information Processing Systems (NeurIPS)},
  year      = {2021}
}

@inproceedings{yin2022avit,
  author    = {Yin, Hongxu and Vahdat, Arash and Alvarez, Jose M. and
               Mallya, Arun and Kautz, Jan and Molchanov, Pavlo},
  title     = {{A-ViT}: Adaptive Tokens for Efficient Vision Transformer},
  booktitle = {Proc. IEEE/CVF Conf. Computer Vision and Pattern Recognition (CVPR)},
  year      = {2022},
  doi       = {10.1109/CVPR52688.2022.01054}
}

@inproceedings{tsai2017graphene,
  author    = {Tsai, Chia-Che and Porter, Donald E. and Vij, Mona},
  title     = {{Graphene-SGX}: A Practical Library {OS} for Unmodified
               Applications on {SGX}},
  booktitle = {Proc. USENIX Annual Technical Conf. (ATC)},
  pages     = {645--658},
  year      = {2017}
}

@inproceedings{paszke2019pytorch,
  author    = {Paszke, Adam and Gross, Sam and Massa, Francisco and Lerer, Adam and
               Bradbury, James and Chanan, Gregory and Killeen, Trevor and
               Lin, Zeming and Gimelshein, Natalia and Antiga, Luca and
               Desmaison, Alban and K{\"o}pf, Andreas and Yang, Edward and
               DeVito, Zachary and Raison, Martin and Tejani, Alykhan and
               Chilamkurthy, Sasank and Steiner, Benoit and Fang, Lu and
               Bai, Junjie and Chintala, Soumith},
  title     = {{PyTorch}: An Imperative Style, High-Performance Deep Learning
               Library},
  booktitle = {Advances in Neural Information Processing Systems (NeurIPS)},
  pages     = {8024--8035},
  year      = {2019}
}

@misc{howard2019imagenette,
  author       = {Howard, Jeremy},
  title        = {Imagenette: A Smaller Subset of 10 Easily Classified Classes
                  from {ImageNet}},
  howpublished = {GitHub repository},
  url          = {https://github.com/fastai/imagenette},
  year         = {2019}
}

@article{lecun1998mnist,
  author    = {LeCun, Yann and Bottou, L{\'e}on and Bengio, Yoshua and
               Haffner, Patrick},
  title     = {Gradient-Based Learning Applied to Document Recognition},
  journal   = {Proceedings of the IEEE},
  volume    = {86},
  number    = {11},
  pages     = {2278--2324},
  year      = {1998},
  doi       = {10.1109/5.726791}
}

@misc{hendrycks2016gelu,
  author       = {Hendrycks, Dan and Gimpel, Kevin},
  title        = {Gaussian Error Linear Units ({GELUs})},
  howpublished = {arXiv:1606.08415},
  year         = {2016}
}

@misc{wolfyosys,
  author       = {Wolf, Claire Xenia},
  title        = {Yosys Open {SYnthesis} Suite},
  year         = {2026},
  url          = {https://yosyshq.net/yosys/},
  note         = {Accessed: Nov.~16, 2026}
}

@misc{hai2025aiindex,
  author = {{Stanford Institute for Human-Centered Artificial Intelligence}},
  title = {The 2025 {AI} Index Report},
  year = {2025},
  url = {https://hai.stanford.edu/ai-index/2025-ai-index-report}
}

@inproceedings{tramer2019slalom,
  author = {Tram{\`e}r, Florian and Boneh, Dan},
  title = {Slalom: Fast, Verifiable and Private Execution of Neural Networks in Trusted Hardware},
  booktitle = {Proc. Int. Conf. Learning Representations (ICLR)},
  year = {2019},
  url = {https://openreview.net/forum?id=rJVorjCcKQ}
}

@misc{hunt2018chiron,
  author = {Hunt, Tyler and Song, Congzheng and Shokri, Reza and Shmatikov, Vitaly and Witchel, Emmett},
  title = {Chiron: Privacy-preserving Machine Learning as a Service},
  howpublished = {arXiv:1803.05961},
  year = {2018},
  url = {https://arxiv.org/abs/1803.05961}
}

@inproceedings{lequoc2020securetf,
  author = {{Le Quoc}, Do and Gregor, Franz and Arnautov, Sergei and Kunkel, Roland and Bhatotia, Pramod and Fetzer, Christof},
  title = {{secureTF}: A Secure {TensorFlow} Framework},
  booktitle = {Proc. Int. Middleware Conf. (Middleware)},
  pages = {44--59},
  year = {2020},
  doi = {10.1145/3423211.3425687},
  url = {https://dl.acm.org/doi/10.1145/3423211.3425687}
}

@inproceedings{zhai2022scaling,
  author = {Zhai, Xiaohua and Kolesnikov, Alexander and Houlsby, Neil and Beyer, Lucas},
  title = {Scaling Vision Transformers},
  booktitle = {Proc. IEEE/CVF Conf. Computer Vision and Pattern Recognition (CVPR)},
  pages = {12104--12113},
  year = {2022},
  url = {https://openaccess.thecvf.com/content/CVPR2022/papers/Zhai_Scaling_Vision_Transformers_CVPR_2022_paper.pdf}
}

@article{kalyanapu2025amxgating,
  author = {Kalyanapu, Joshua and Dizani, Farshad and Ghanbari, Azam and Asher, Darsh and {Mirbagher Ajorpaz}, Samira},
  title = {Exploiting {Intel AMX} Power Gating},
  journal = {IEEE Computer Architecture Letters},
  volume = {24},
  number = {1},
  pages = {113--116},
  year = {2025},
  doi = {10.1109/LCA.2025.3555183},
  url = {https://research.ece.ncsu.edu/brainspec/wp-content/uploads/sites/35/2026/06/exploiting_intel_amx_power_gating.pdf}
}

@inproceedings{xie2024osworld,
  author = {Xie, Tianbao and Zhang, Danyang and Chen, Jixuan and Li, Xiaochuan and Zhao, Siheng and Cao, Ruisheng and {Toh Jing Hua} and Cheng, Zhoujun and Shin, Dongchan and Lei, Fangyu and Liu, Yitao and Xu, Yiheng and Zhou, Shuyan and Savarese, Silvio and Xiong, Caiming and Zhong, Victor and Yu, Tao},
  title = {{OSWorld}: Benchmarking Multimodal Agents for Open-Ended Tasks in Real Computer Environments},
  booktitle = {Advances in Neural Information Processing Systems (NeurIPS), Datasets and Benchmarks Track},
  volume = {37},
  year = {2024},
  doi = {10.52202/079017-1650}
}

@inproceedings{liu2023llava,
  author    = {Liu, Haotian and Li, Chunyuan and Wu, Qingyang and Lee, Yong Jae},
  title     = {Visual Instruction Tuning},
  booktitle = {Advances in Neural Information Processing Systems (NeurIPS)},
  year      = {2023}
}

@article{fedus2022switch,
  author    = {Fedus, William and Zoph, Barret and Shazeer, Noam},
  title     = {Switch Transformers: Scaling to Trillion Parameter Models with
               Simple and Efficient Sparsity},
  journal   = {Journal of Machine Learning Research},
  volume    = {23},
  number    = {120},
  pages     = {1--39},
  year      = {2022}
}

@inproceedings{shokri2017membership,
  author    = {Shokri, Reza and Stronati, Marco and Song, Congzheng and Shmatikov, Vitaly},
  title     = {Membership Inference Attacks Against Machine Learning Models},
  booktitle = {Proc. IEEE Symp. Security and Privacy (S\&P)},
  pages     = {3--18},
  year      = {2017}
}

@inproceedings{fredrikson2014pharmacogenetics,
  author    = {Fredrikson, Matthew and Lantz, Eric and Jha, Somesh and Lin, Simon and Page, David and Ristenpart, Thomas},
  title     = {Privacy in Pharmacogenetics: An End-to-End Case Study of Personalized Warfarin Dosing},
  booktitle = {Proc. USENIX Security Symp.},
  pages     = {17--32},
  year      = {2014}
}

@inproceedings{mehnaz2022sensitive,
  author    = {Mehnaz, Shagufta and Dibbo, Sayanton V. and Kabir, Ehsanul and Li, Ninghui and Bertino, Elisa},
  title     = {Are Your Sensitive Attributes Private? Novel Model Inversion Attribute Inference Attacks on Classification Models},
  booktitle = {Proc. USENIX Security Symp.},
  pages     = {4579--4596},
  year      = {2022}
}

@misc{intel2020rapl,
  author    = {{Intel Corporation}},
  title     = {Running Average Power Limit Energy Reporting / {CVE-2020-8694}, {CVE-2020-8695}},
  year      = {2020},
  url       = {https://www.intel.com/content/www/us/en/developer/articles/technical/software-security-guidance/advisory-guidance/running-average-power-limit-energy-reporting.html},
  note      = {Accessed: Nov.~16, 2026}
}

@misc{intel2022throttling,
  author    = {{Intel Corporation}},
  title     = {Frequency Throttling Side Channel Guidance},
  year      = {2022},
  url       = {https://www.intel.com/content/www/us/en/developer/articles/technical/software-security-guidance/best-practices/frequency-throttling-side-channel-guidance.html},
  note      = {Accessed: Nov.~16, 2026}
}

@misc{intel2022doit,
  author    = {{Intel Corporation}},
  title     = {Data Operand Independent Timing {ISA} Guidance},
  year      = {2022},
  url       = {https://www.intel.com/content/www/us/en/developer/articles/technical/software-security-guidance/best-practices/data-operand-independent-timing-isa-guidance.html},
  note      = {Accessed: Nov.~16, 2026}
}

@misc{intel2022consttime,
  author    = {{Intel Corporation}},
  title     = {Guidelines for Mitigating Timing Side Channels Against Cryptographic Implementations},
  year      = {2022},
  url       = {https://www.intel.com/content/www/us/en/developer/articles/technical/software-security-guidance/secure-coding/mitigate-timing-side-channel-crypto-implementation.html},
  note      = {Accessed: Nov.~16, 2026}
}

@inproceedings{chen2026uncorebleed,
  author    = {Decheng Chen and Zhi Zhang and Zhenkai Zhang and Xin Zhang and Yansong Gao and Yi Zou},
  title     = {{UncoreBleed}: {AEX-Free}, {High-Resolution}, and {Low-Noise} {Side-Channel} Attacks on {SGX} Enclaved Execution},
  booktitle = {Proc. USENIX Security Symp.},
  year      = {2026},
  pages     = {5355--5371},
  publisher = {USENIX Association},
  url       = {https://www.usenix.org/conference/usenixsecurity26/presentation/chen-decheng}
}

@misc{costan2016sgx,
  author       = {Costan, Victor and Devadas, Srinivas},
  title        = {{Intel SGX} Explained},
  howpublished = {Cryptology ePrint Archive, Paper 2016/086},
  year         = {2016}
}

@inproceedings{teerapittayanon2016branchynet,
  author    = {Teerapittayanon, Surat and McDanel, Bradley and Kung, H. T.},
  title     = {{BranchyNet}: Fast Inference via Early Exiting from Deep
               Neural Networks},
  booktitle = {Proc. Int. Conf. Pattern Recognition (ICPR)},
  pages     = {2464--2469},
  year      = {2016}
}

\end{document}